\documentclass[fleqn,twoside]{article}
\usepackage{amsmath,amssymb,espcrc2,color,cite}

\usepackage{graphicx,soul}
\usepackage[figuresright]{rotating}

\newcommand{\url}[1]{{\tt #1}}
\newcommand{\lsim}
{\;\raisebox{-.3em}{$\stackrel{\displaystyle <}{\sim}$}\;}
\newcommand{\gsim}
{\;\raisebox{-.3em}{$\stackrel{\displaystyle >}{\sim}$}\;}

\def\order#1{\ensuremath{{\cal O}(#1)}}
\newcommand{\gmt}{\ensuremath{(g-2)_\mu}}
\newcommand{\br}{{\rm BR}}
\newcommand{\bsg}{BR($b \to s \gamma$)}

\newcommand{\bmm}{\ensuremath{\br(B_s \to \mu^+\mu^-)}}
\newcommand{\ssi}{\ensuremath{\sigma^{\rm SI}_p}}

\newcommand{\Mh}{M_h}

\newcommand{\MA}{M_A}

\newcommand{\mgl}{m_{\tilde g}}

\newcommand{\msqr}{m_{\tilde{q}_R}}

\newcommand{\mneu}[1]{m_{\tilde \chi^0_{#1}}}

\newcommand{\mstaue}{m_{\staue}}
\newcommand{\staue}{\tilde \tau_1}

\newcommand{\tb}{\tan\beta}

\newcommand{\tev}{\,\, \mathrm{TeV}}
\newcommand{\gev}{\,\, \mathrm{GeV}}

\definecolor{Orange}{named}{Orange}
\definecolor{Purple}{named}{Purple}

\newcommand{\ETslash}{/ \hspace{-.7em} E_T}

\graphicspath{{figs/}}

\title{\bf Higgs and Supersymmetry
 \\ \vspace{0.5em}}

\author{
{\bf O.~Buchmueller}\address[Imperial]
   {High\,Energy\,Physics\,Group,\,Blackett\,Laboratory,\,Imperial\,College,\,Prince\,Consort\,Road,\,London\,SW7\,2AZ,\,UK},
{\bf R.~Cavanaugh}\address[FNAL]
   {Fermi National Accelerator Laboratory, P.O. Box 500, 
    Batavia, Illinois 60510, USA}\hbox{$^{\rm ,}$}\address[UIC]
   {Physics Department, University of Illinois at Chicago, Chicago, 
    Illinois 60607-7059, USA},
{\bf A.~De Roeck}\address[CERN]
   {CERN, CH--1211 Gen\`eve 23, Switzerland}\hbox{$^{\rm ,}$}\address[Antwerpen]
   {Antwerp University, B--2610 Wilrijk, Belgium},
 {\bf M.J.~Dolan}\address[IPPP]
{Institute for Particle Physics
     Phenomenology,\,University\,of\,Durham,\,South 
     Road,\,Durham\,DH1\,3LE,\,UK},
{\bf J.R.~Ellis}\address[KCL]{Theoretical Particle Physics
  and Cosmology Group, Department of Physics, King's College London, London~WC2R~2LS, UK}\hbox{$^{\rm ,}$}\addressmark[CERN], 
{\bf H.~Fl\"acher}\address[Rochester]
   {H.H.~Wills Physics Laboratory, University of Bristol, Tyndall Avenue, Bristol BS8 1TL, UK},
{\bf S.~Heinemeyer}\address[Santander]
   {Instituto de F\'{\i}sica de Cantabria (CSIC-UC), 
    E--39005 Santander, Spain},
{\bf G.~Isidori}\address[Frascati]
   {INFN, Laboratori Nazionali di Frascati, Via E. Fermi 40, 
    I--00044 Frascati, Italy},
{\bf J.~Marrouche}\addressmark[Imperial],
   {\bf D.~Mart\'inez~Santos}\addressmark[CERN],
{\bf K.A.~Olive}\address[Minnesota] 
{William I.\ Fine Theoretical Physics Institute, School of Physics and
 Astronomy, University of Minnesota, Minneapolis, Minnesota 55455, USA}, 
{\bf S.~Rogerson}\addressmark[Imperial],
{\bf F.J.~Ronga}\address[ETHZ]
   {Institute for Particle Physics, ETH Z\"urich, CH--8093 Z\"urich, 
   Switzerland},
{\bf K.J.~de~Vries}\addressmark[Imperial],
{\bf G.~Weiglein}\address[DESY]
   {DESY, Notkestrasse 85, D--22607 Hamburg, Germany}
}

\begin{document}

\begin{abstract}
Global frequentist fits to the CMSSM and NUHM1 using the
{\tt MasterCode} framework predicted $\Mh \simeq 119 \gev$ in fits
incorporating the \gmt\ constraint and $\simeq 126 \gev$ without it.
Recent results by ATLAS and CMS could be compatible with a Standard
Model-like Higgs boson around $\Mh \simeq 125 \gev$. 
We use the previous 
{\tt MasterCode} analysis to calculate the likelihood for a measurement of 
any nominal Higgs mass within the range of 115 to 130~GeV. Assuming a
Higgs mass measurement at $\Mh \simeq 125 \gev$,  
we display updated global likelihood contours in the $(m_0, m_{1/2})$
and other parameter planes of the CMSSM and NUHM1, and present updated
likelihood functions for $\mgl, \msqr$, \bmm\ and the spin-independent
dark matter cross section \ssi. The implications of dropping \gmt\ from the
fits are also discussed. 
We furthermore comment on a hypothetical measurement of
$\Mh \simeq 119 \gev$.

\bigskip
\begin{center}
{\tt KCL-PH-TH/2011-40, LCTS/2011-21, CERN-PH-TH/2011-305, \\
DCPT/11/168, DESY 11-242, IPPP/11/84, FTPI-MINN-11/31, UMN-TH-3023/11, 
{arXiv:1112.3564 [hep-ph]}}
\end{center}
\vspace{2.0cm}
\end{abstract}

\maketitle

\section{Introduction}
\label{sec:intro}

Taking into account the relevant experimental constraints, the CMSSM and
NUHM1 predict that the lightest Higgs boson should have couplings
similar to those of the Standard Model (SM) Higgs
boson~\cite{Ellis:2001qv,Ambrosanio:2001xb,mc7}, and that it 
should weigh no more than $\sim 130 \gev$~\cite{mh,FeynHiggs,asbstev}. We
recently reported the results of global frequentist fits within the CMSSM and
NUHM1 to the first $\sim 1$/fb of LHC data, also including precision
electroweak and flavour measurements and the XENON100 upper limit on
elastic spin-independent dark matter scattering~\cite{mc7}, updating the
results of previous global fits by
ourselves~\cite{mc1,mc2,mc3,mc35,mc4,mc5,mc6,mcweb} and
others~\cite{pre-LHC,post-LHC} (see also \cite{LR}). The results
reported in~\cite{mc7} 
included likelihood contours in the $(m_0, m_{1/2})$, $(\tan \beta,
m_{1/2})$ and $(\MA, \tan \beta)$ planes of the CMSSM and NUHM1, as well
as $\Delta \chi^2$ functions for $\mgl$, \bmm, $\Mh, \MA$ and sparticle
production thresholds in $e^+ e^-$ annihilation. 

Notable predictions of
these global fits included $\Mh = 119.1^{+3.4}_{-2.9} \gev$ in the CMSSM
and $\Mh = 118.8^{+2.7}_{-1.1} \gev$ in the NUHM1 (which should be combined
with an estimated theory error $\Delta \Mh = \pm 1.5 \gev$). 
These two fits are based solely on the Higgs-{\it independent} searches
including the 
\gmt\ constraint, i.e., they do not 
rely on the existing limits from LEP~\cite{Barate:2003sz,Schael:2006cr},
the Tevatron~\cite{TevHiggs}, or the LHC~\cite{ATLASHA,CMSHA}.
These predictions increase to $\Mh = 124.8^{+3.4}_{-10.5} \gev$ in the CMSSM
and $126.6^{+0.7}_{-1.9} \gev$ in the NUHM1 if the \gmt\ constraint is dropped.

Subsequently, the ATLAS and CMS Collaborations have released their
official combination of the searches for a SM Higgs boson with the first
$\sim 1$/fb of LHC luminosity at $E_{\rm cm} =
7$~TeV~\cite{ATLAS+CMS}. Impressively, the combination excludes a SM
Higgs boson with a mass between 141 and 476~GeV. Most recently, the ATLAS and
CMS Collaborations have presented preliminary updates of their results with
$\sim 5$/fb of data~\cite{Dec13}. These results may be compatible with a SM-like Higgs
boson around $\Mh \simeq 125 \gev$, though CMS also report an
excess at $\Mh \simeq 119 \gev$ in the $ZZ^*$ channel.
We recall that, for low values of $\Mh$, the SM electroweak vacuum would be unstable~\cite{unstable}, 
decaying into a state with Higgs vev $> 10^8 (10^{10}) \gev$ if $\Mh = 119 (125) \gev$,
and that a very plausible mechanism
for stabilizing the vacuum is supersymmetry (SUSY)~\cite{ER}.

In this paper, we first report the likelihood function for an LHC
measurement of $\Mh$ with a nominal value $\in (115, 130) \gev$,
incorporating the theoretical error $\pm 1.5 \gev$ and an estimate 
$\pm 1 \gev$ of the possible experimental error. 
In both the CMSSM and NUHM1, this likelihood function is minimized for
$\Mh \simeq 119 \gev$ if \gmt\ is included, and is contained within the
theoretical uncertainty range shown previously as a `red band'~\cite{mc7}.  We
then discuss the consequences of combining 
a measurement of $\Mh \simeq 125 \gev$ (assuming that the current
excess will be confirmed with more integrated luminosity) with our previous
analysis~\cite{mc7} of constraints on the CMSSM and NUHM1 including \gmt.

We find that the best-fit values of $m_0$ and $m_{1/2}$ in the CMSSM and NUHM1
are moved to 
substantially higher values, especially in the case of $m_{1/2}$. 
We also update our results 
on the best-fit regions in the $(m_{1/2}, \tb)$ and $(\MA, \tb)$ planes,
where we find again the substantial increase in $m_{1/2}$, 
as compared with our pre-LHC $\Mh$ results.
We present the corresponding one-dimensional likelihood functions for
the gluino mass $\mgl$, an average right-handed squark mass $\msqr$, the
lighter scalar tau mass, $\mstaue$,
as well as in the $(\mneu{1}, \ssi)$ plane, where $\mneu{1}$ is the mass of
the lightest neutralino and \ssi\ is the spin-independent dark matter
scattering cross section. As could be expected, we find larger values of
$\mgl, \msqr, \mneu{1}$ and $\mstaue$ than in our pre-LHC $\Mh$ fit, and smaller values of
\ssi, though \bmm\ is little affected.

Since $\Mh \simeq 125 \gev$ is the value that was favoured in the
CMSSM/NUHM1 fits omitting the \gmt\ constraint~\cite{mc7}, we also show some
results for fits where \gmt\ is dropped. In this case, we find that
preferred regions of the $(m_0, m_{1/2})$ planes are localized at
relatively high values, corresponding to relatively large sparticle
masses. Correspondingly, the spin-independent dark matter scattering
cross section \ssi\ would be relatively small in this case, though again there
would be relatively little effect on \bmm.

Finally, we show selected results for a hypothetical measurement of 
$\Mh \simeq 119 \gev$.


\section{Prediction for \boldmath{$\Mh$}}

We recall that the independent parameters of the CMSSM \cite{cmssm1} may be
taken as 
the common values of the scalar and fermionic supersymmetry-breaking
masses $m_0, m_{1/2}$ at the GUT scale, the supposedly
universal trilinear soft supersymmetry-breaking parameter, $A_0$,
and the ratio of Higgs v.e.v.'s, $\tb$. A study of the distribution of Higgs
masses in the CMSSM was performed in~\cite{enos}. Motivated by \gmt\ and, to a lesser
extent, \bsg, we assume that the Higgs mixing parameter $\mu > 0$.
In the case of the NUHM1~\cite{nuhm1}, we relax the universality assumption
for the soft supersymmetry-breaking contributions to the two Higgs masses,
$m_{H_u}^2 = m_{H_d}^2$, allowing $m^2_{H_u} = m^2_{H_d} \neq m^2_0$. 

In our previous
papers~\cite{mc1,mc2,mc3,mc35,mc4,mc5,mc6,mc7} we constructed a global
likelihood function that receives contributions from electroweak
precision observables, B-decay measurements, the XENON100 direct search
for dark matter scattering~\cite{XE100} and the LHC searches for supersymmetric
signals, calculated within the  {\tt MasterCode} framework~\cite{mcweb}.  
This incorporates code based on~\cite{Svenetal} as well as {\tt
  SoftSUSY}~\cite{Allanach:2001kg}, {\tt FeynHiggs 2.8.6}~\cite{FeynHiggs},
{\tt SuFla}~\cite{SuFla}, {\tt SuperIso}~\cite{SuperIso}, 
{\tt MicrOMEGAs}~\cite{MicroMegas} and {\tt SSARD}~\cite{SSARD}, using
the SUSY Les Houches Accord~\cite{SLHA}. As before, we use a Markov
Chain Monte Carlo (MCMC) approach to sample the parameter spaces of
supersymmetric models, and the results of this paper are based on the 
sample of 70M CMSSM points and another 125M NUHM1 points, both
extending up to $m_0, m_{1/2} = 4000 \gev$. 

We used in~\cite{mc7} the public results of searches for supersymmetric 
signals using $\sim 1$/fb of LHC data analyzed by the ATLAS and CMS 
Collaborations and $\sim 0.3$/fb of data analyzed by the LHCb Collaboration.
These include searches for jets + $\ETslash$ events without leptons by
ATLAS~\cite{ATLASsusy} and CMS~\cite{CMSsusy},  
searches for the heavier MSSM Higgs bosons, $H/A$~\cite{ATLASHA,CMSHA},
and new upper limits on \bmm\ from the CMS~\cite{CMSbmm},
LHCb~\cite{LHCbbmm} and CDF Collaborations~\cite{CDFbmm}. 
Our global frequentist fit~\cite{mc7} yielded regions of the CMSSM and NUHM1
parameter spaces that are preferred at the 68 and 95\% CL. 

This was the
basis in~\cite{mc7} for the predictions
$\Mh = 119.1^{+3.4}_{-2.9} \gev$ in the CMSSM and 
$\Mh = 118.8^{+2.7}_{-1.1} \gev$ in the NUHM1, if the \gmt\ constraint is included 
as calculated using the {\tt FeynHiggs} code 
which is quoted as having a theoretical error $\pm 1.5 \gev$~\cite{FeynHiggs}. 
It is important to note that these best-fit
values are well above the LEP lower limit and below the Tevatron/LHC upper
limit on $\Mh$, which played no role in their determination. Fig.~12
of~\cite{mc7} displayed the $\Delta \chi^2$ likelihood functions for the 
{\tt FeynHiggs} value of $\Mh$ in these models as blue lines, 
with the theoretical error $\pm 1.5 \gev$ represented by
red bands in these plots. As already noted, these predictions increase to
$\Mh = 124.8^{+3.4}_{-10.5} \gev$ in the CMSSM
and $126.6^{+0.7}_{-1.9} \gev$ in the NUHM1 if the \gmt\ constraint is dropped.
The uncertainty on the $\Mh$ prediction is somewhat asymmetric, which is
due to the different constraints coming into play. At low $\Mh$ values, the most important
constraint is that due to the LHC, though other low-energy constraints also play roles.
On the other hand, at high values of $\Mh$, it rises logarithmically with the scalar top masses, so
increasing $\Mh$ increases exponentially the required supersymmetry-breaking mass scales, and worsens the agreement
with other low-energy data and the CDM constraint.

\subsection*{Results without a Higgs-boson mass measurement}

\noindent\\
Within the supersymmetric frameworks discussed here, a confirmation of the
excess reported by ATLAS and CMS~\cite{Dec13} and consequently the discovery
of a SM-like Higgs boson is expected 
to be possible in the coming year, with a mass 
in the range 
between 114 and 130~GeV~\cite{Dec13}. We assume that this
measurement will 
yield a nominal value of $\Mh$ within this range, with an experimental
error that we estimate as $\pm 1\gev$. We now estimate the
one-dimensional likelihood function for the nominal central value of $\Mh$,
which may be written as 
$\Mh = \Mh^{\tt FH} + \Delta \Mh^{\rm Th} + \Delta \Mh^{\rm Exp}$,
where $\Mh^{\tt FH}$ denotes the output of {\tt FeynHiggs} (which was plotted
in Fig.~12 of~\cite{mc7} for the fits including \gmt),
$\Delta \Mh^{\rm Th}$ denotes its difference from the true value of $\Mh$ (the
theoretical error estimated as $\pm 1.5 \gev$), and $\Delta \Mh^{\rm Exp}$
denotes the experimental error in measuring $\Mh$ (estimated as $\pm 1 \gev$).
Here we treat the experimental and the theoretical errors as Gaussians, and
include them as supplementary uncertainties in the fit for the nominal central
value of $\Mh$. As a consequence of including these uncertainties,
the $\Delta \chi^2$ function for the nominal central value of $\Mh$ presented
here differs slightly from the $\Delta \chi^2$ function for the 
{\tt FeynHiggs} estimate  
$\Mh^{\tt FH}$ shown in Fig.~12 of~\cite{mc7}.

\begin{figure*}[htb!]
\resizebox{8.3cm}{!}{\includegraphics{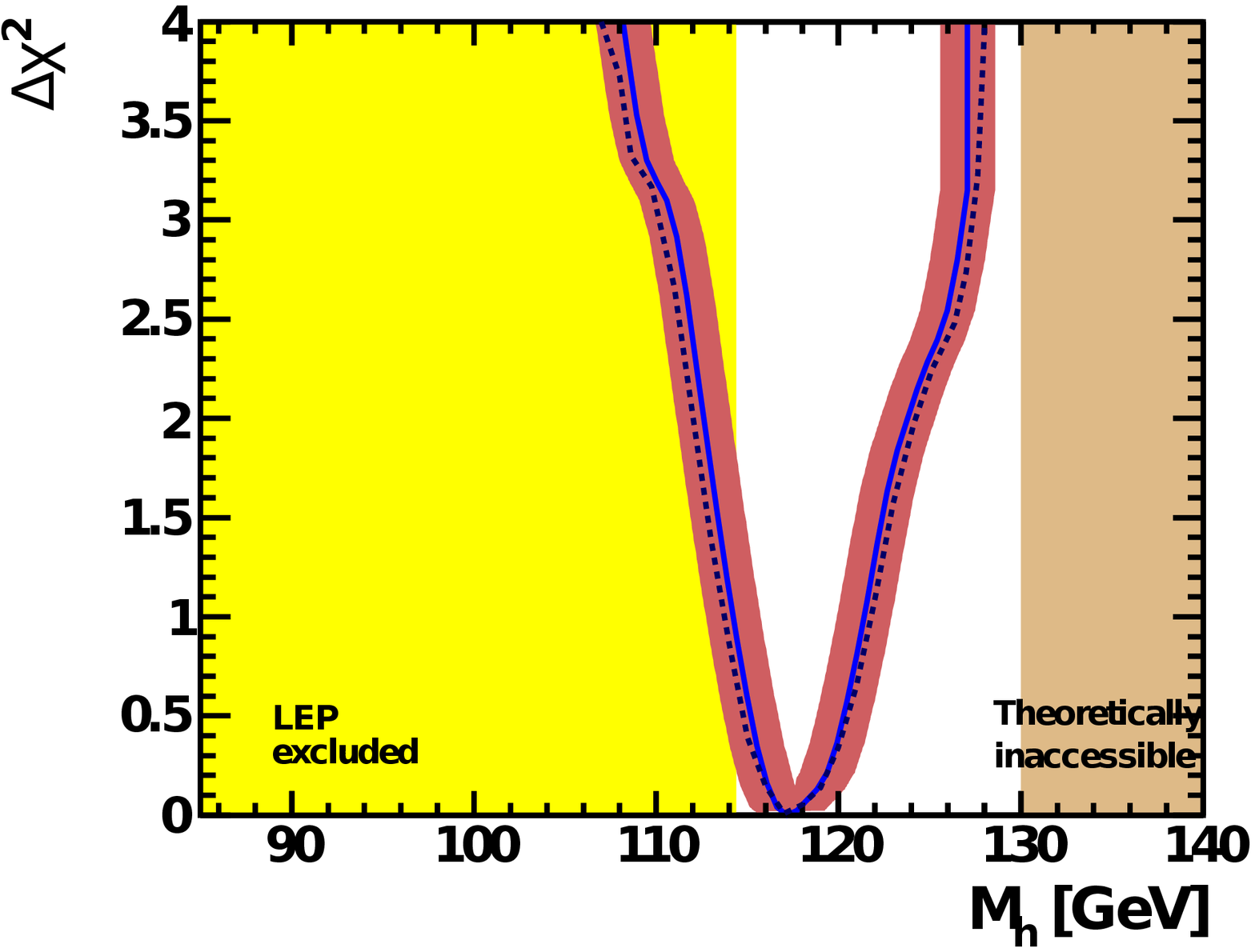}}
\resizebox{8.3cm}{!}{\includegraphics{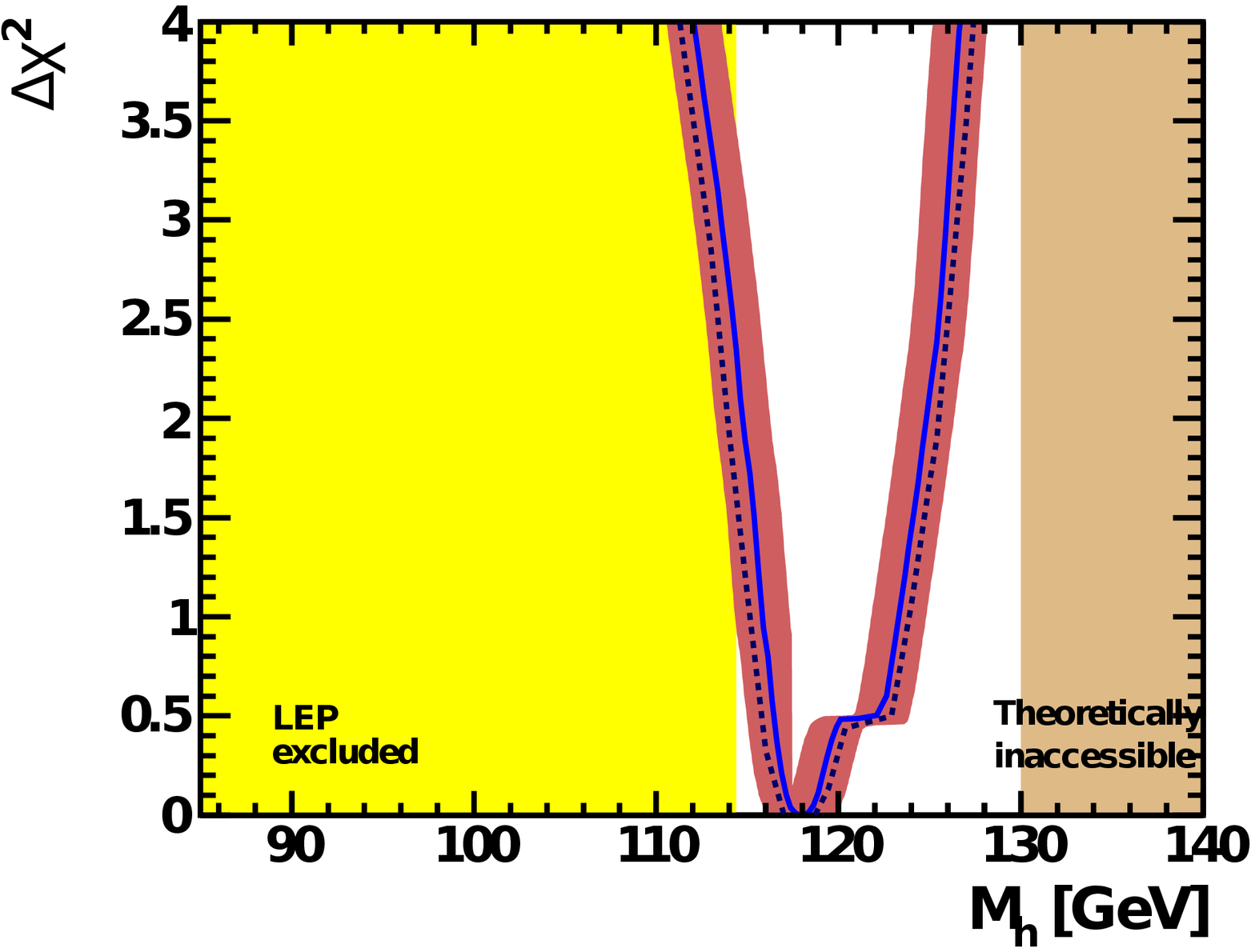}}
\vspace{-1cm}
\caption{\it 
The one-dimensional $\Delta \chi^2$ functions for $\Mh$ in
the CMSSM (left) and the NUHM1 (right). The solid lines are for
fits including all the available data including \gmt\ but excluding the direct
LEP~\protect\cite{Barate:2003sz,Schael:2006cr},
Tevatron~\protect\cite{TevHiggs} and earlier LHC~\protect\cite{ATLASHA,CMSHA} 
constraints on $\Mh$, with a red band indicating the estimated theoretical
uncertainty in the calculation of $\Mh$
of $\sim 1.5 \gev$. The dashed line shows the
$\Delta \chi^2$ likelihood function for the nominal
central value of a hypothetical LHC measurement of $\Mh$, as
estimated on the basis of the frequentist analysis of~\protect\cite{mc7},
and allowing for an experimental error of $\pm 1 \gev$ in the measurement of
$\Mh$ and a theoretical error of $\pm 1.5 \gev$ in the {\tt FeynHiggs}
calculation of $\Mh$ at any given point in the parameter space.
}
\label{fig:likelyMh}
\end{figure*}

We see in Fig.~\ref{fig:likelyMh}~%
\footnote{The `theoretically inaccessible' area of $\Mh > 130 \gev$ could in
  principle be extended to higher values if one extended the scanned ranges of $m_{1/2}$ and
  $m_0$, which are both restricted here to below $\sim 4 \tev$, as discussed above. 
  However, due to the logarithmic dependence of $\Mh$, one would gain
  only about one GeV even if values up to $10 \tev$ were included.}%
~that the values of $\Delta \chi^2$ for the
nominal 
value of $\Mh$ calculated in the CMSSM and NUHM1  with the \gmt\ constraint
and including both the theoretical and experimental errors lie below the blue
lines taken from Fig.~12 of~\cite{mc7}. This is to be expected, since the
calculation of the dashed line incorporates additional uncertainties. 
As is also to be expected, in each case the calculated $\Delta \chi^2$
lies within the previous red band. The
most likely nominal value of the LHC measurement of $\Mh$ remains
$\Mh \simeq 119 \gev$ in both the CMSSM and NUHM1. 
A value of $\Mh \simeq 125 \gev$
is disfavoured in our analysis by $\Delta \chi^2 = 2.2$ in the
CMSSM and by 1.6 in the NUHM1 if \gmt\ is included. For comparison, a nominal
value of  $\Mh = 114 \gev$, corresponding roughly to the lower limit set by LEP
for an SM-like Higgs boson~\cite{Barate:2003sz,Schael:2006cr}, 
has $\Delta \chi^2 = 0.8 (1.5)$.
On the other hand, if we drop \gmt\ there is essentially no $\chi^2$ 
price to be paid by including a measurement of $\Mh \simeq 125 \gev$.


\section{Implementation of the LHC Constraint on \boldmath{$\Mh$}}

We now study the possibility that the LHC experiments confirm the excess reported around
$125 \gev$ and indeed discover a SM-like Higgs boson. Assuming
\begin{align}
\Mh = 125 \pm 1 ({\rm exp.}) \pm 1.5 ({\rm theo.}) \gev~,
\label{Mh125}
\end{align} 
we incorporate this new constraint using the `afterburner' approach 
discussed previously~\cite{mc7} for other observables. 
This value would be favoured if \gmt\ were
dropped from our global CMSSM or NUHM1 fit~\cite{mc7}. Alternatively, a
measurement of such a high $\Mh$ value could point to the realization of some
different (possibly GUT-based) version of the MSSM 
(see, for instance, \cite{AbdusSalam:2011fc}).
We also mention briefly some implications if $\Mh \simeq  119 \gev$.


\subsubsection*{Comments on the LHC data}

\noindent\\
As a preamble to these studies, we first comment on the results of the
current ATLAS/CMS Higgs combination. 
We recall that the {\it local} $p$-value for the background-only hypothesis
for the excess found in the ATLAS data at $\Mh \simeq 125 \gev$ is $p = 1.9 \times 10^{-4}$,
while that in the CMS data at $\Mh \simeq 125 \gev$ has $p = 5 \times 10^{-3}$. In addition,
CMS reports an excess in the $ZZ^*$ channel at $\Mh = 119 \gev$ with similar
significance, but this is not confirmed by ATLAS.

In order to assess the {\it global} $p$-value of a potential
signal, one should take the `look-elsewhere effect' (LEE) into
account. This is conventionally estimated by 
adding to the {\it local} $p$-value the quantity 
$N \times$ exp($-Z_{\rm max}^2/2$), where $N$ is the number of times the
observed upper limit on the signal crosses over the $\mu  = \sigma/\sigma_{\rm SM} = 0$ level in
the upward direction, and $Z_{\rm max}$ is the maximal signal
significance~\cite{Dec13}. Accounting for the LEE,
ATLAS assess the {\it global} $p$-value of their excess at $125 \gev$ In the range $(110, 146) \gev$
to be 0.6\%, and CMS assess the significance of their excess at $125 \gev$ to be
1.9\% in the range $(110, 145) \gev$. 
 
On the other hand, as the CMSSM and NUHM1 naturally require 
$\Mh \lsim 130 \gev$,
the LEE factor is strongly reduced in these frameworks. 

Since the excess around $125 \gev$ is common to both experiments
and has the correct signal 
strength: $\mu \approx 1$ can be interpreted as a 
Higgs signal in either the SM or a
supersymmetric framework. We focus here on this interpretation,
commenting subsequently on some implications if $\Mh \simeq 119 \gev$.

\begin{figure*}[htb!]
\resizebox{8.6cm}{!}{\includegraphics{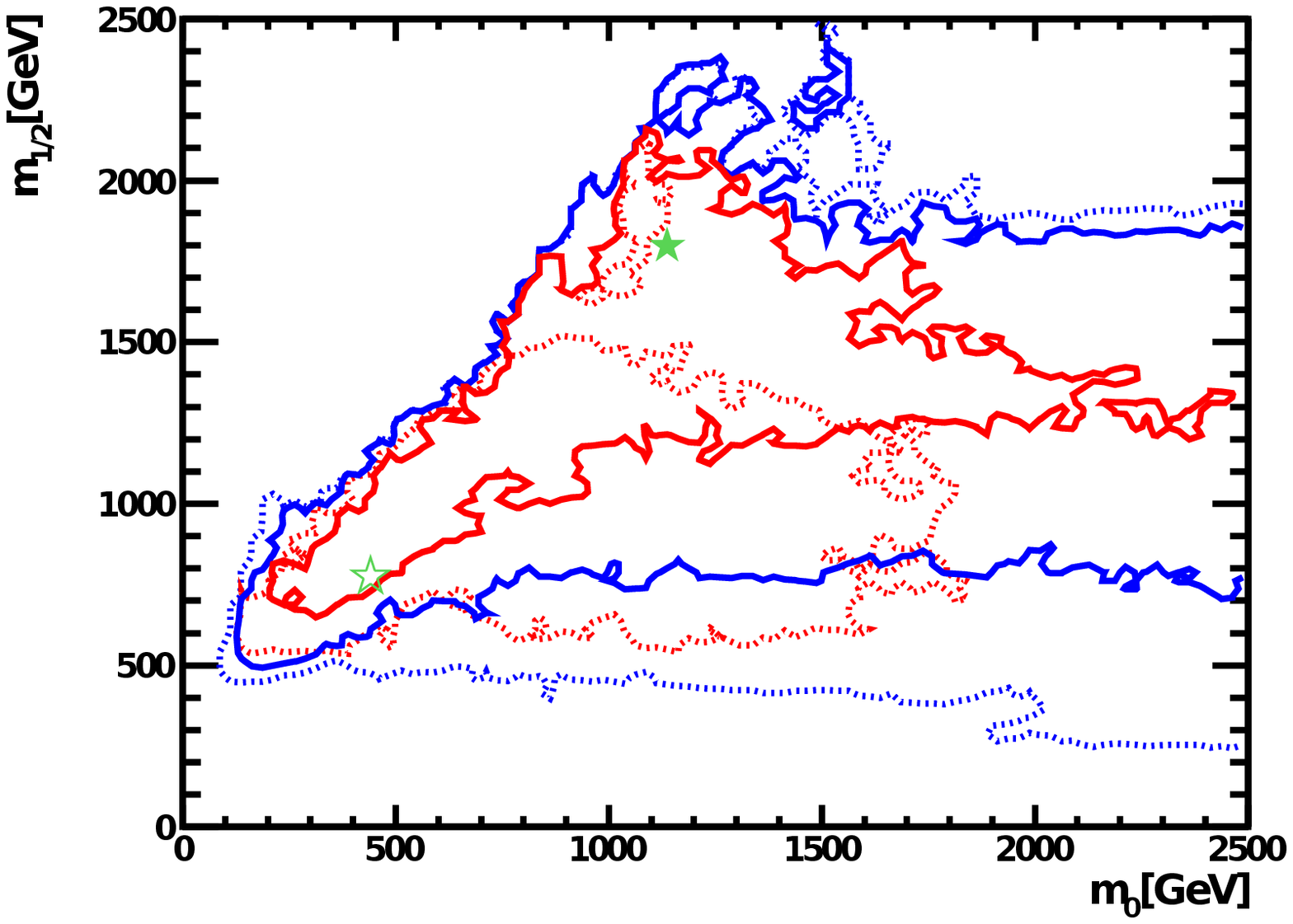}}
\resizebox{8.6cm}{!}{\includegraphics{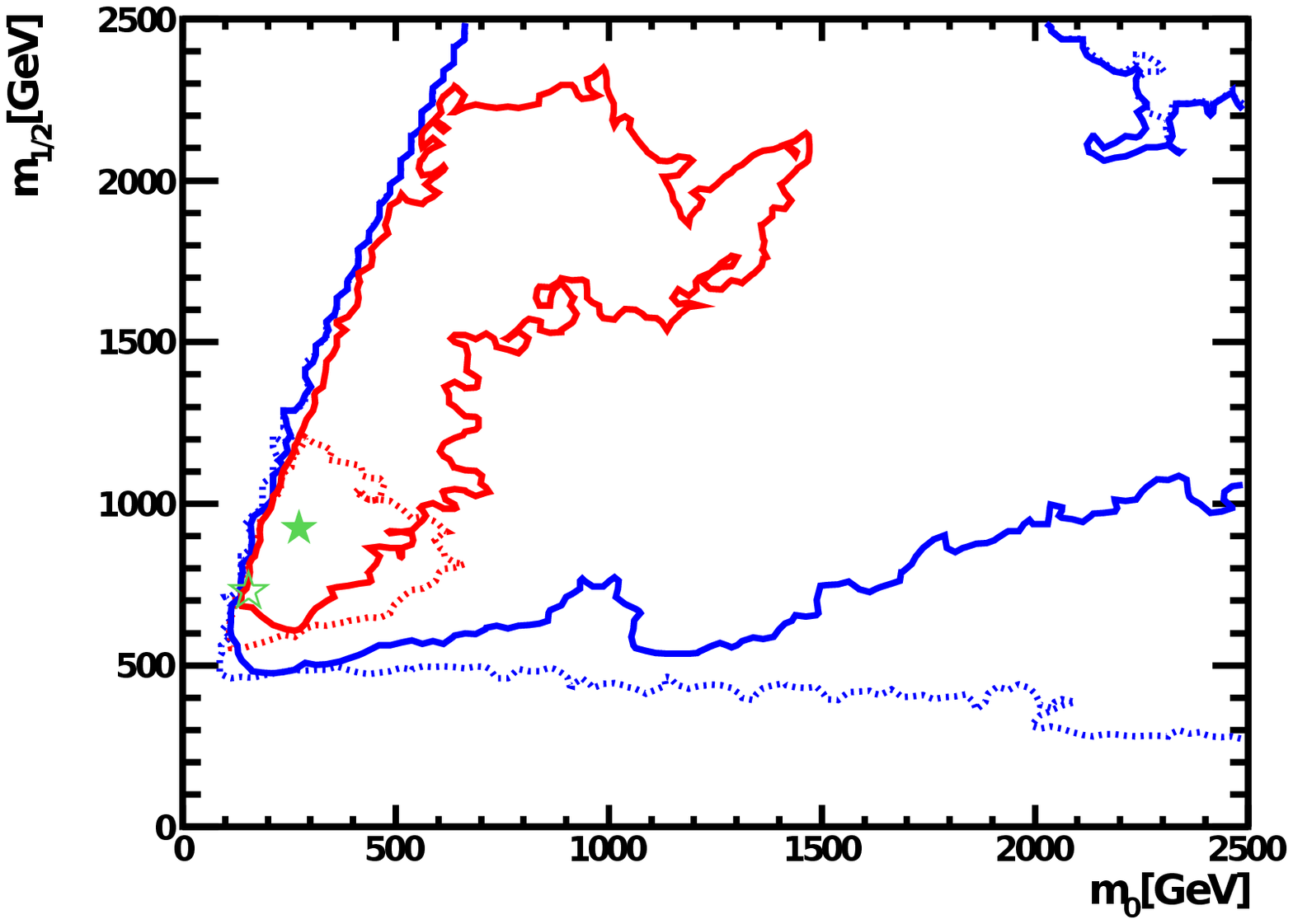}}
\caption{\it The $(m_0, m_{1/2})$ planes in the CMSSM (left) and the
  NUHM1 (right). The $\Delta \chi^2= 2.30$ and $5.99$   contours,
  commonly interpreted as the boundaries of the 68 and 95\% CL regions,
  are indicated in red and blue, respectively, the solid lines including
  the hypothetical LHC measurement $\Mh = 125 \pm 1 \gev$
  and allowing for a theoretical error $\pm 1.5 \gev$,
  and the dotted lines showing the contours
  found previously in~\protect\cite{mc7} without this $\Mh$ constraint.
  Here the open green stars denote the pre-Higgs best-fit
  points~\protect\cite{mc7}, whereas the solid green stars indicate the new
  best-fit points.
}
\label{fig:6895}
\end{figure*}

\subsubsection*{What if \boldmath{$\Mh = 125 \gev$}?}

\noindent\\
We first examine the effects on
the global likelihood functions in various CMSSM and NUHM1
parameter planes, and then study implications for various observables of a potential LHC measurement
$\Mh \simeq 125 \gev$, see Eq.~(\ref{Mh125}).
The $(m_0, m_{1/2})$ planes
shown in Fig.~\ref{fig:6895} are for the CMSSM (left) and NUHM1 (right)~\footnote{We have omitted 
from the NUHM1 sample displayed here and in subsequent figures a grouping of points at large
$m_{1/2}$ and small $m_0$ for which different codes yield discrepant values of the relic density.
{\tt MicrOMEGAs}~\cite{MicroMegas} and {\tt DarkSusy}~\cite{DarkSusy} yield densities within the WMAP range for these points,
whereas {\tt SuperIso Relic}~\cite{SuperIsoRelic} and {\tt SSARD}~\cite{SSARD} both yield substantially lower densities. The other figures shown
in this paper are not affected significantly by the omission of these points (which have $\Delta \chi^2 > 5$), 
pending resolution of this discrepancy. Tests in other
regions of the NUHM1 sample have not revealed significant discrepancies between these codes.}.
The regions preferred at the 68\%~CL are outlined in red, and those favoured at the 
95\%~CL are outlined in blue. The solid (dotted) lines include (omit) the assumed LHC
Higgs constraint.
The open green star denotes the pre-Higgs best-fit point~\cite{mc7},
whereas the solid green star indicates the new best-fit point incorporating
a Higgs-boson mass measurement at $125 \gev$.

Since in the CMSSM and NUHM1 the radiative corrections 
contributing to the value of $\Mh$ are sensitive primarily to $m_{1/2}$
and $\tb$, and only to a lesser extent to $m_0$ 
(the stop masses, which are the most relevant for $\Mh$, depend mostly on
$m_{1/2}$ due to the RGE running, and only mildly on $m_0$)%
, we expect that the primary
effect of imposing the $\Mh$ constraint should be to affect the
preferred ranges of $m_{1/2}$ and $\tb$, with a lesser effect on the preferred 
range of $m_0$.
This effect is indeed seen in both panels of Fig.~\ref{fig:6895}. 
We see that the 68\% CL ranges of $m_{1/2}$ extend to
somewhat larger
values and with a wider spread than the pre-Higgs results, particularly in the NUHM1.
However, the NUHM1 best-fit value of $m_{1/2}$ remains at a relatively low value
of $\sim 800 \gev$, whereas the best-fit value of $m_{1/2}$ in the CMSSM moves to
$\sim 1900 \gev$. This jump reflects the flatness of the likelihood function
for $m_{1/2}$ between $\sim 700 \gev$ and $\sim 2 \tev$, which
is also reflected later in the one-dimensional $\Delta \chi^2$ functions for some
sparticle masses~\footnote{Our fits are relatively insensitive to $A_0$, so we do not
display figures for this parameter.}.

When we add the hypothetical $\Mh$ constraint 
the total $\chi^2$ at the best-fit points
increases substantially, as seen in Table~\ref{tab:p-comp}, and the $p$-value
decreases correspondingly. 
The Table compares fit probabilities for two different 
assumptions on the Higgs boson mass measurements $\simeq 119, 125 \gev$, see
above, and with the option of dropping the \gmt\ constraint in the latter case~%
\footnote{The fit
probabilities are indicative of the current experimental data
preferences for one scenario over another but, as discussed in~\cite{mc7}, but
they do not provide a robust confidence-level estimation for the
actual choice made by Nature.}.
The combination of the increase in $\chi^2$ and in the increase in the number of d.o.f. leads to a
substantially lower $p$-value after the inclusion of Eq.~(\ref{Mh125}), if \gmt\ is taken into account. 
On the other hand, a hypothetical
mass measurement at  $119 \gev$ 
would yield an improvement in the fit.
For comparison, we also show the parameters for the best-fit
points. Since the uncertainties are large and highly non-Gaussian, we omit
them from the Table.

\begin{table*}[!tbh!]
\renewcommand{\arraystretch}{1.5}
\begin{center}
\begin{tabular}{|c||c|c|c|c|c|c|} \hline
Model & Minimum & Fit Prob- & $m_{1/2}$ & $m_0$ & $A_0$ & $\tb$ \\
      & $\chi^2$/d.o.f.& ability & (GeV) & (GeV) & (GeV) & \\
\hline \hline
CMSSM  & & & & & &  \\
\hline
pre-Higgs
                     & 28.8/22 & 15\% & 780 & 440 & $-1120$ & 40 \\
$\Mh \simeq 125 \gev$, \gmt & 31.0/23 & 12\% & 1800 & 1140 & $ 1370$ & 46 \\
$\Mh \simeq 125 \gev$, no \gmt & 21.3/22 & 50\% & 1830 & 1320 & $ -1860$ & 47 \\
$\Mh \simeq 119 \gev$& 28.9/23 & 18\% &  880 &  400 & $-890$ & 38 \\

\hline\hline
NUHM1 & & & & & &  \\
\hline
pre-Higgs     
                     & 26.9/21 & 17\% &  730 &  150  & $-910$ & 41 \\
$\Mh \simeq 125 \gev$, \gmt & 28.9/22 & 15\% & 920 & 270 & $ 1730$ & 27 \\
$\Mh \simeq 125 \gev$, no \gmt & 19.7/21 & 52\% & 2060 & 1400 & $ 2610$ & 46 \\
$\Mh \simeq 119 \gev$ & 27.1/22 & 20\% &  750 &  150 & $ -420$ & 34 \\
\hline
\end{tabular}
\caption{\it 
Comparison of the best-fit points found in the CMSSM and NUHM1
  pre-Higgs~\protect\cite{mc7} and for the two potential LHC Higgs mass
  measurements discussed in the text: $\Mh \simeq 119$ and $125 \gev$.
  In the latter case, we also quote results if the \gmt\ constraint is dropped.
  At the best-fit NUHM1 points, the common values of the soft supersymmetry-breaking
  contributions to the Higgs squared masses are the following - pre-Higgs: $-1.2 \times 10^6 \gev^2$,
  with $\Mh \simeq 125 \gev$ and \gmt: $-5.5 \times 10^6 \gev^2$, 
  with $\Mh \simeq 125 \gev$ but without \gmt: $-8.6 \times 10^5 \gev^2$,
  with $\Mh \simeq 119 \gev$ and \gmt: $-1.2 \times 10^6 \gev^2$.
}
\label{tab:p-comp}
\end{center}
\end{table*}

\begin{figure*}[htb!]
\resizebox{8.6cm}{!}{\includegraphics{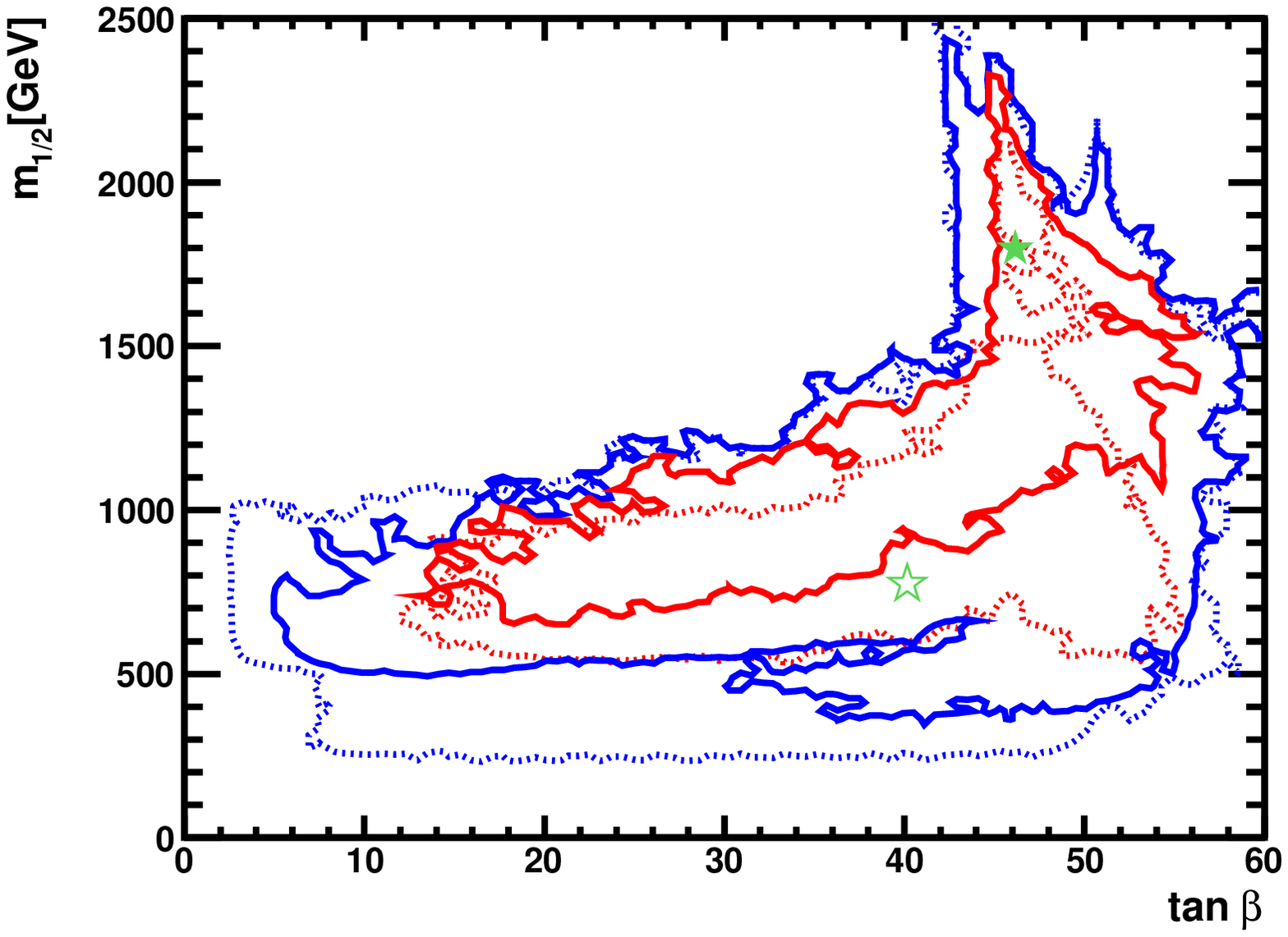}}
\resizebox{8.6cm}{!}{\includegraphics{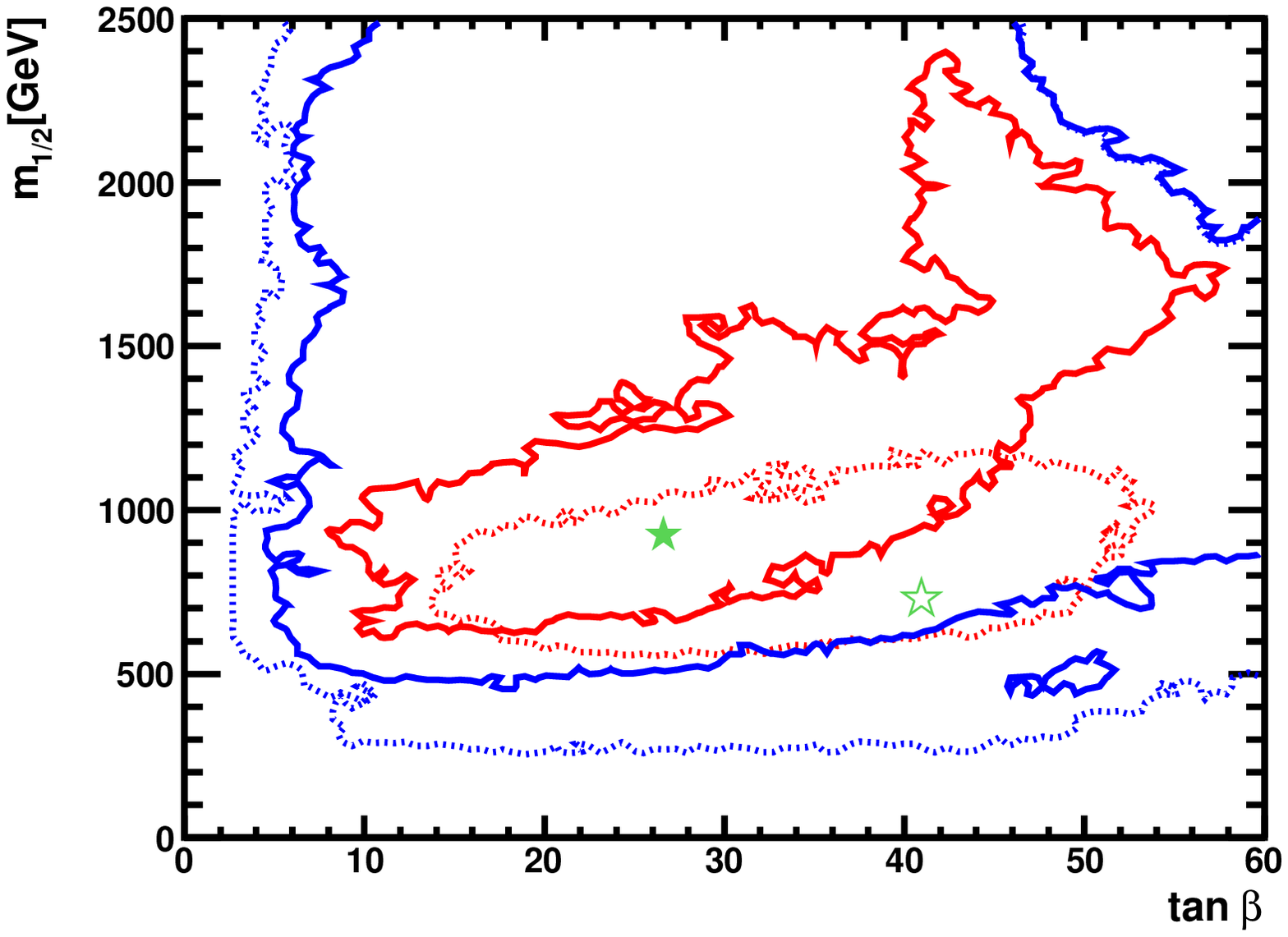}}
\vspace{-1cm}
\caption{\it The $(m_{1/2}, \tb)$ planes in the CMSSM (left) and the
  NUHM1 (right), for $\Mh \simeq 125 \gev$. The notations and
  significations of the contours are 
  the same as in Fig.~\protect\ref{fig:6895}.
}
\label{fig:tanbm12}
\end{figure*}

The restrictions that the hypothetical LHC $\Mh$ constraint imposes on
$m_{1/2}$ are also visible in Fig.~\ref{fig:tanbm12}, where we display
the effects of an LHC $\Mh$ constraint in the $(m_{1/2}, \tb)$ planes of
the CMSSM  and NUHM1. We see here that an LHC $\Mh$ constraint enlarges
visibly the 68\% CL range of $\tb$ in the NUHM1, whereas 
the change is less pronounced in the CMSSM.

\begin{figure*}[htb!]
\resizebox{8.6cm}{!}{\includegraphics{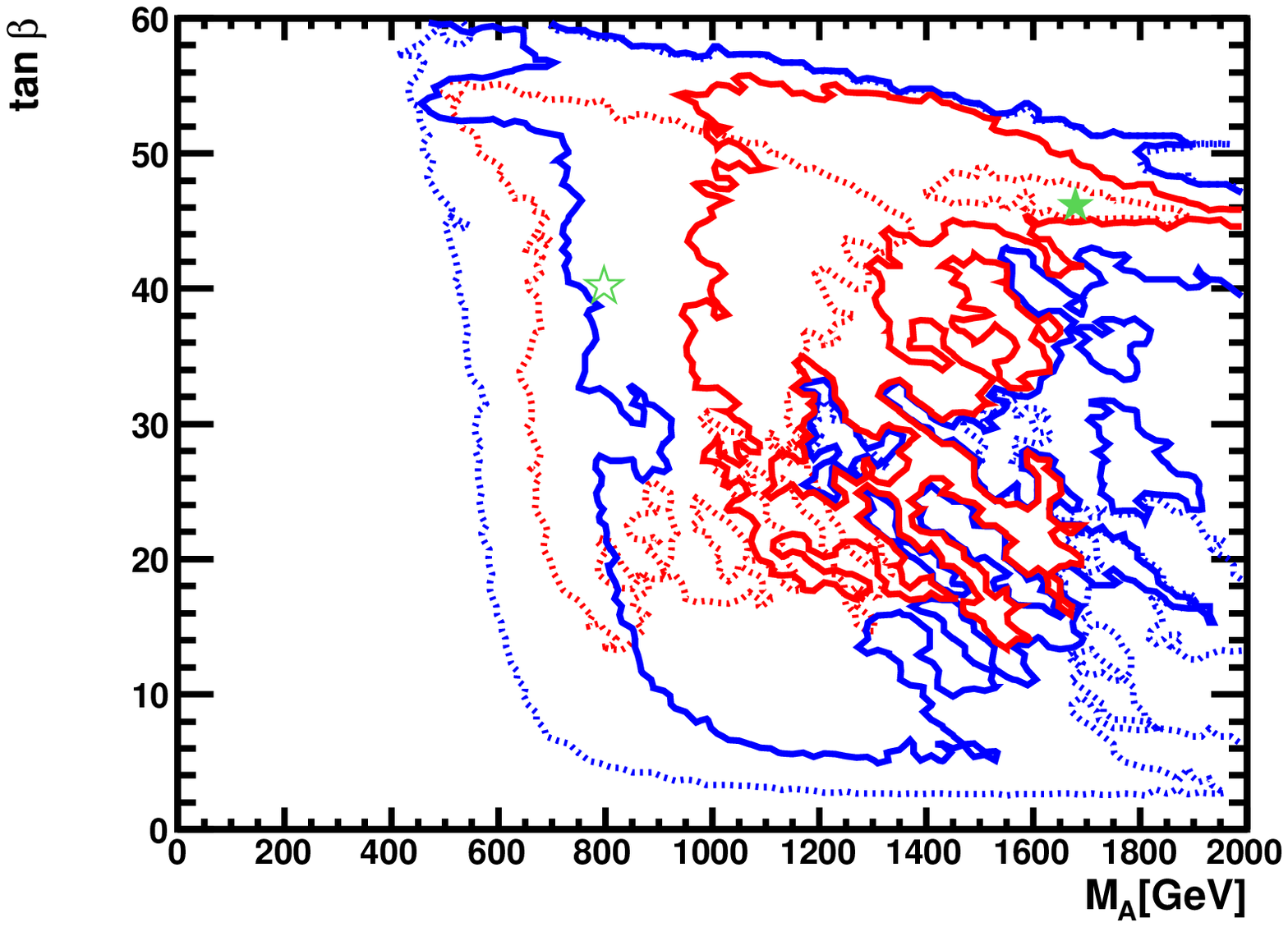}}
\resizebox{8.6cm}{!}{\includegraphics{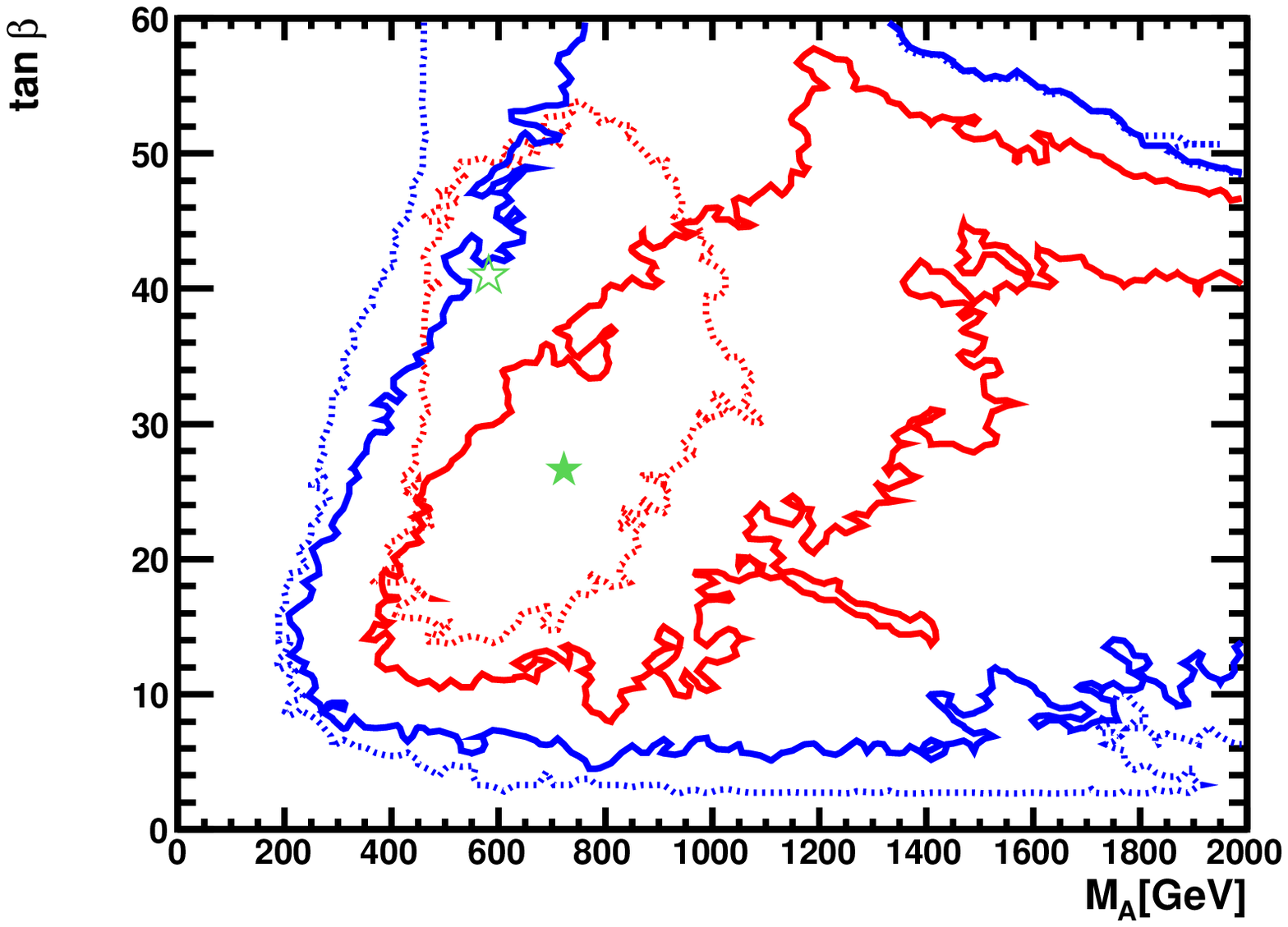}}
\vspace{-1cm}
\caption{\it The $(\MA, \tb)$ planes in the CMSSM (left) and the
  NUHM1 (right), for $\Mh \simeq 125 \gev$. The notations and
  significations of the contours are 
  the same as in Fig.~\protect\ref{fig:6895}.
}
\label{fig:MAtb}
\end{figure*}

The results for the $(\MA, \tb)$ planes in the CMSSM and the NUHM1
are shown in Fig.~\ref{fig:MAtb}. We observe a strong increase in the
best-fit value of $\MA$ in both models, especially in the CMSSM, where
now $\MA \sim 1600 \gev$ is preferred. We re-emphasize, however, that
the likelihood function varies relatively slowly in both models, as compared to the pre-LHC fits.

We now discuss the CMSSM and NUHM1 predictions for some of the
most interesting supersymmetric observables for the LHC in light
of a possible LHC measurement at $\Mh \simeq 125 \gev$.

\begin{figure*}[htb!]
\resizebox{8.5cm}{!}{\includegraphics{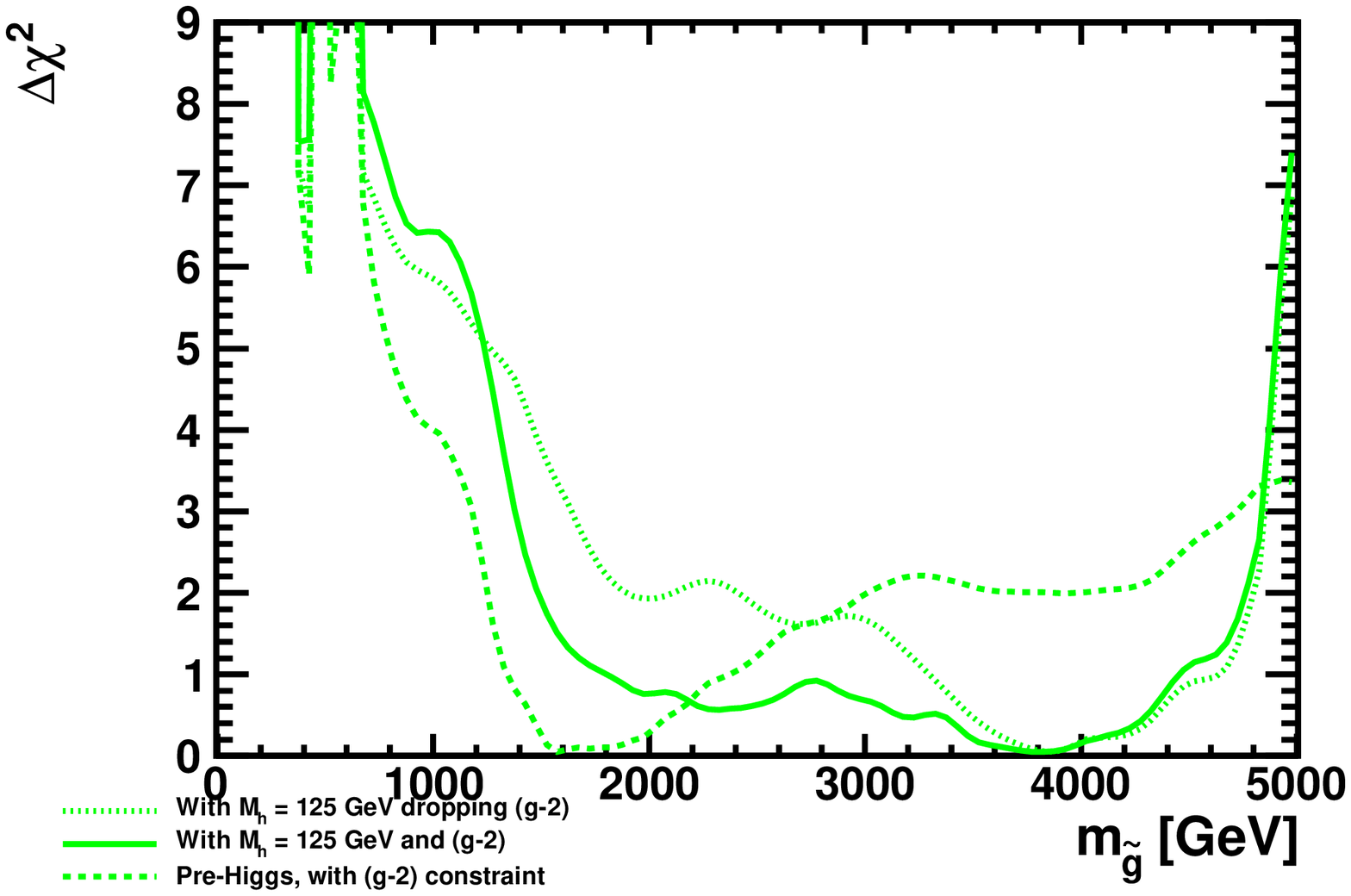}}
\resizebox{8.5cm}{!}{\includegraphics{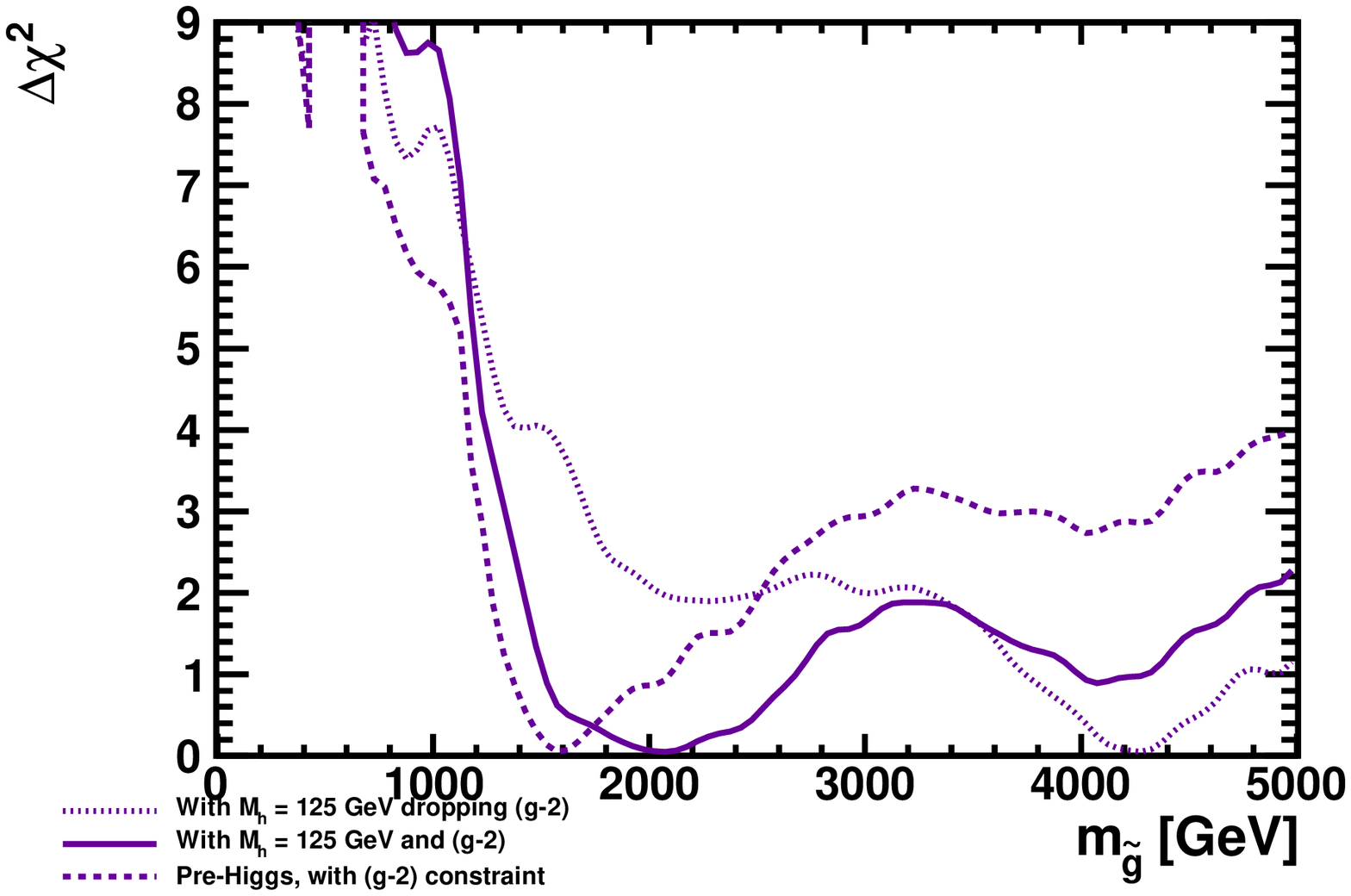}}
\resizebox{8.5cm}{!}{\includegraphics{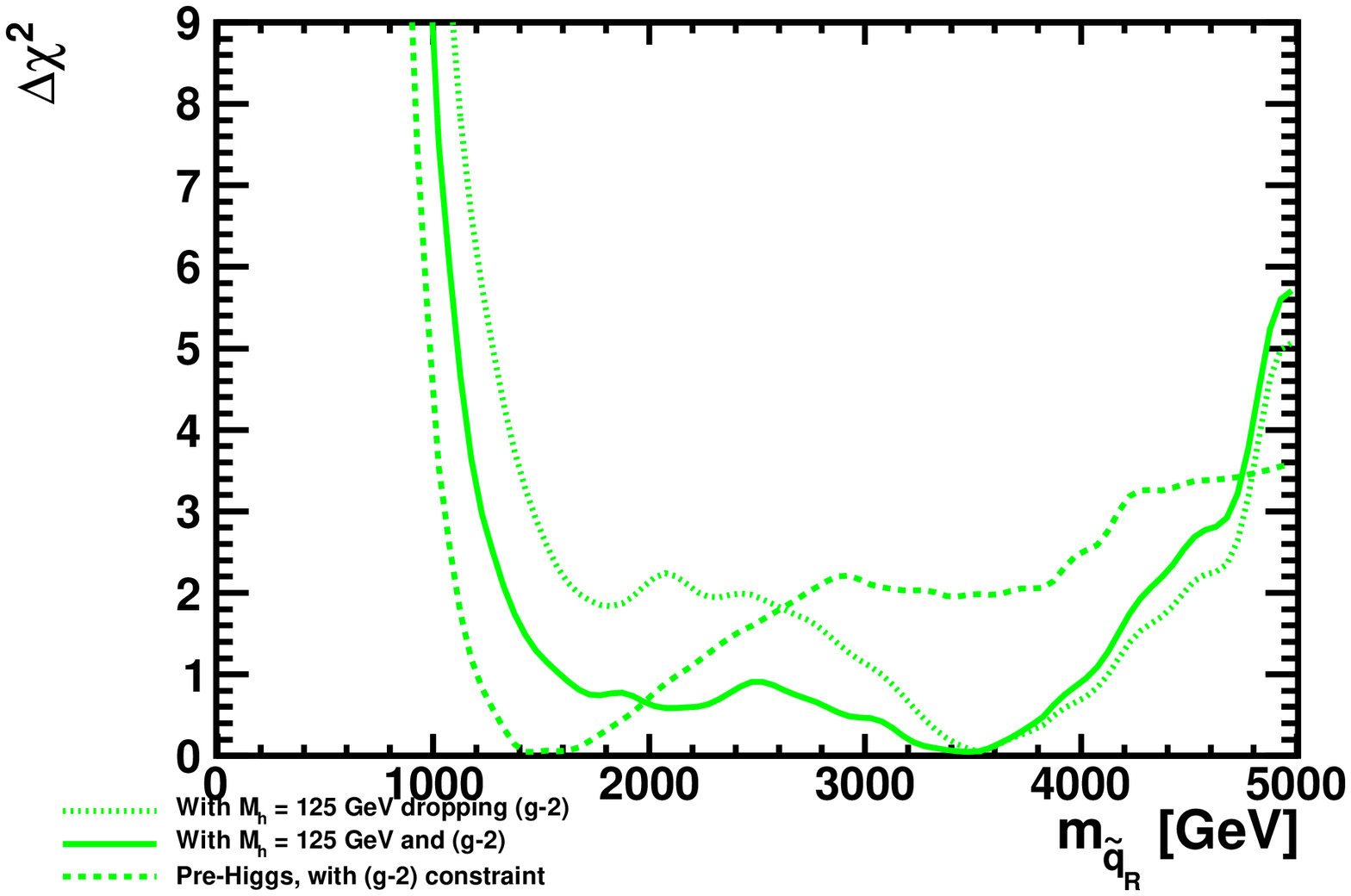}}
\resizebox{8.5cm}{!}{\includegraphics{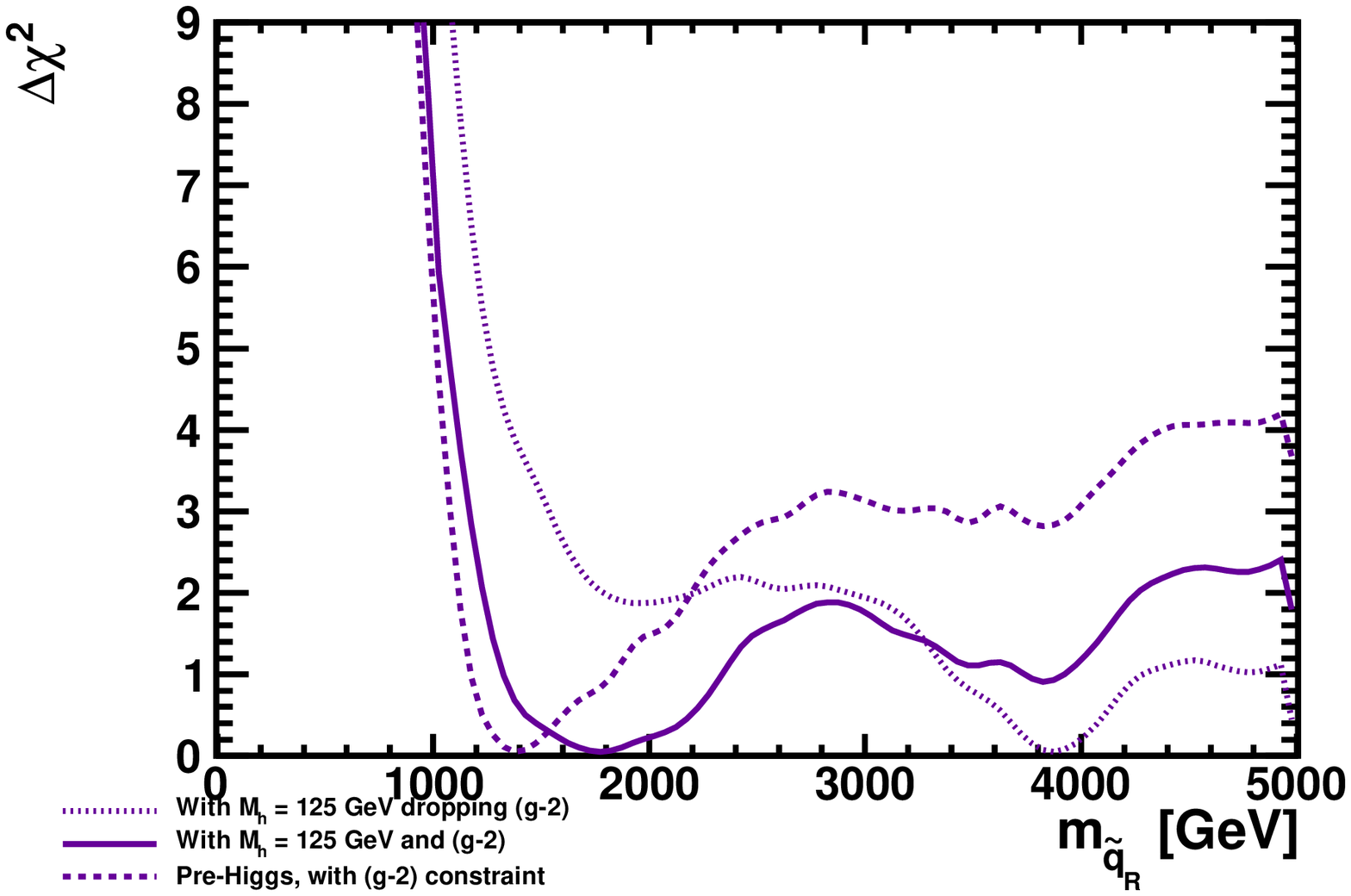}}
\vspace{-1cm}
\caption{\it The one-dimensional $\Delta \chi^2$ functions for $\mgl$ (upper) and
$\msqr$ (lower) in
the CMSSM (left) and the NUHM1 (right). The solid lines are for
fits assuming $\Mh \simeq 125 \gev$ and including \gmt, and the dotted
lines for fits with $\Mh \simeq 125 \gev$ but without \gmt.
The dashed lines show the results for fits without $\Mh \simeq 125 \gev$ but including \gmt~\protect\cite{mc7}.
}
\label{fig:mgl}
\end{figure*}

\begin{figure*}[htb!]
\resizebox{8.2cm}{!}{\includegraphics{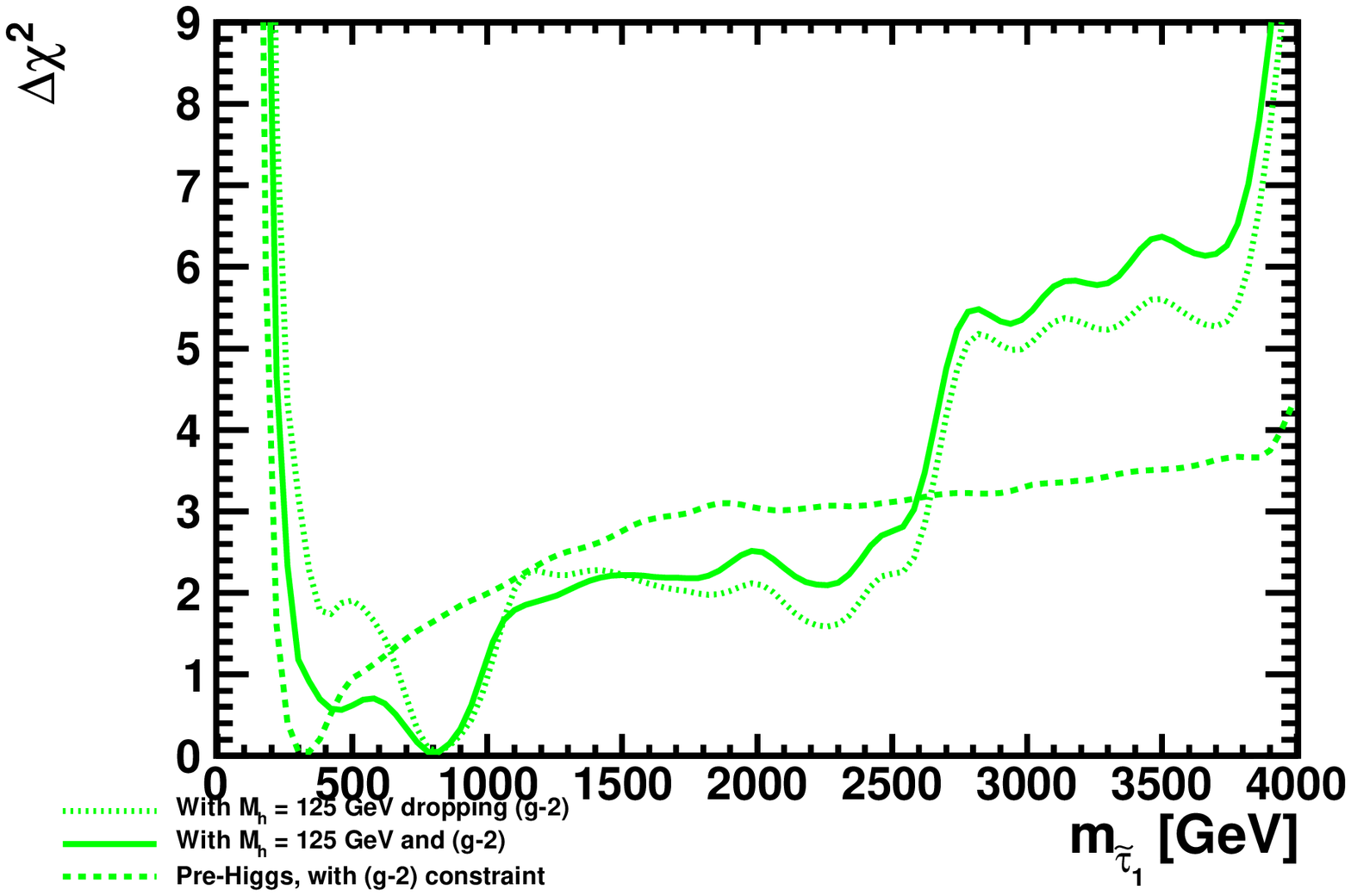}}
\resizebox{8.2cm}{!}{\includegraphics{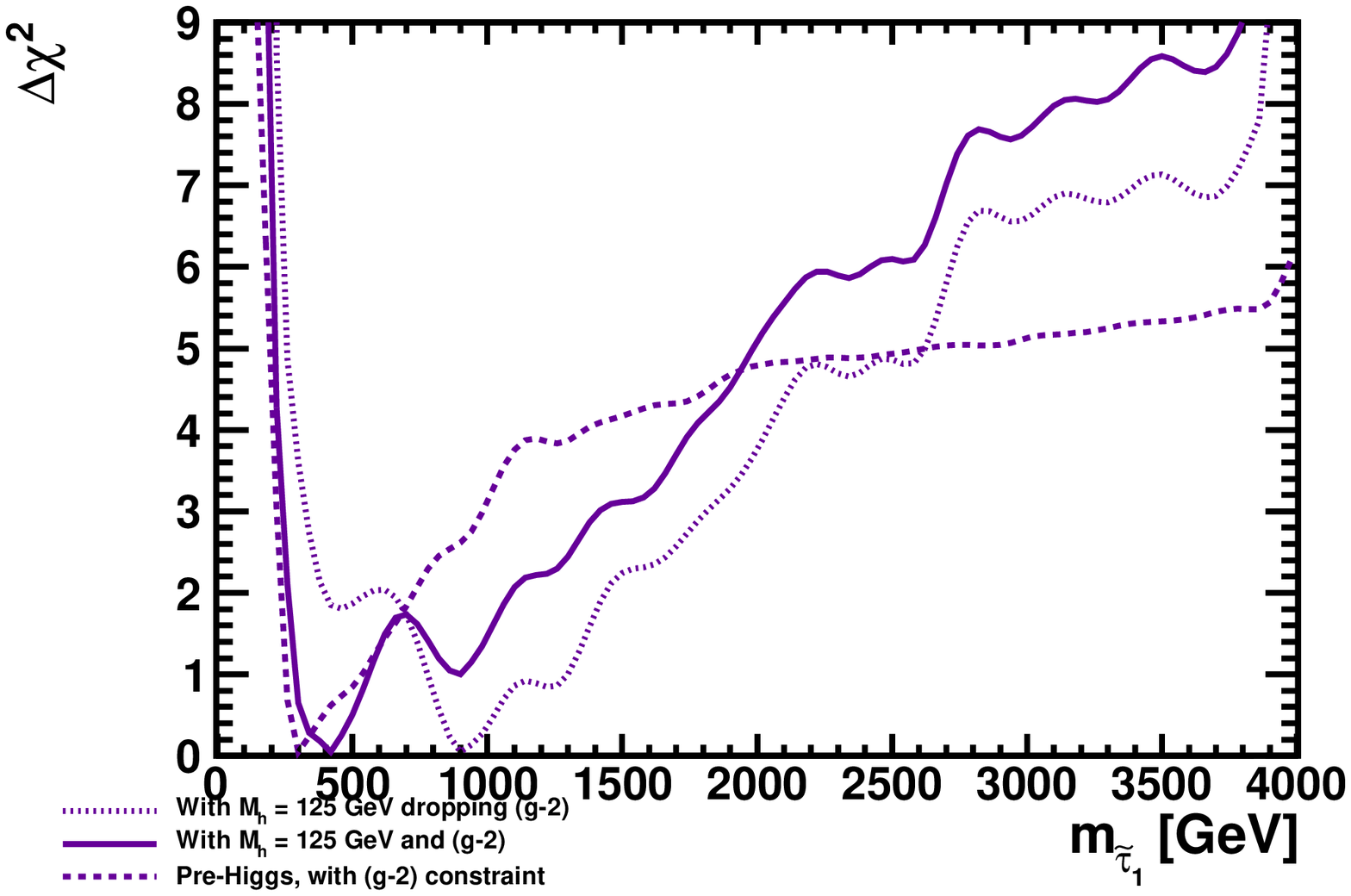}}
\vspace{-1cm}
\caption{\it The one-dimensional $\Delta \chi^2$ functions for the lighter
scalar tau mass $\mstaue$ in
the CMSSM (left) and the NUHM1 (right), for $\Mh \simeq 125 \gev$. 
The notations and significations of the lines are
the same as in Fig.~\protect\ref{fig:mgl}. 
}
\label{fig:mstau}
\end{figure*}

The upper panels of Fig.~\ref{fig:mgl} display the one-dimensional
$\Delta \chi^2$ functions for $\mgl$ before and after applying the new LHC 
$\Mh \simeq 125 \gev$ constraint (dashed and solid lines, respectively, in both cases including \gmt). 
We also show as dotted lines the $\Delta \chi^2$ functions for a fit including $\Mh \simeq 125 \gev$
and dropping \gmt.
As expected on the basis of Fig.~\ref{fig:6895}, the preferred values 
$\mgl \sim 4 \tev$ in the CMSSM are much higher than in our pre-LHC fit and what would be preferred if 
$\Mh \simeq 119 \gev$, and presumably beyond the reach of the LHC.
On the other hand, in the NUHM1 $\mgl \sim 2 \tev$ is marginally preferred.
However, in both models the $\Delta \chi^2$ function varies little over the range $(2, 4) \tev$.
Similar features are found for $\msqr$, as shown in the lower panels of
Fig.~\ref{fig:mgl}. 
In both models, the regions of $\mgl$ and $\msqr$ with $\Delta \chi^2 \lsim 1$
start at masses around 
$1.5 \tev$, leaving a large range accessible to the SUSY searches at the 
LHC. In the case of the lighter stau mass $\mstaue$ for 
$\Mh \simeq 125 \gev$
shown in Fig.~\ref{fig:mstau}, 
we again see preferred masses larger than in the pre-Higgs fit, with favoured
values extending up to $\mstaue \sim 1 \tev$.

\begin{figure*}[htb!]
\resizebox{8.2cm}{!}{\includegraphics{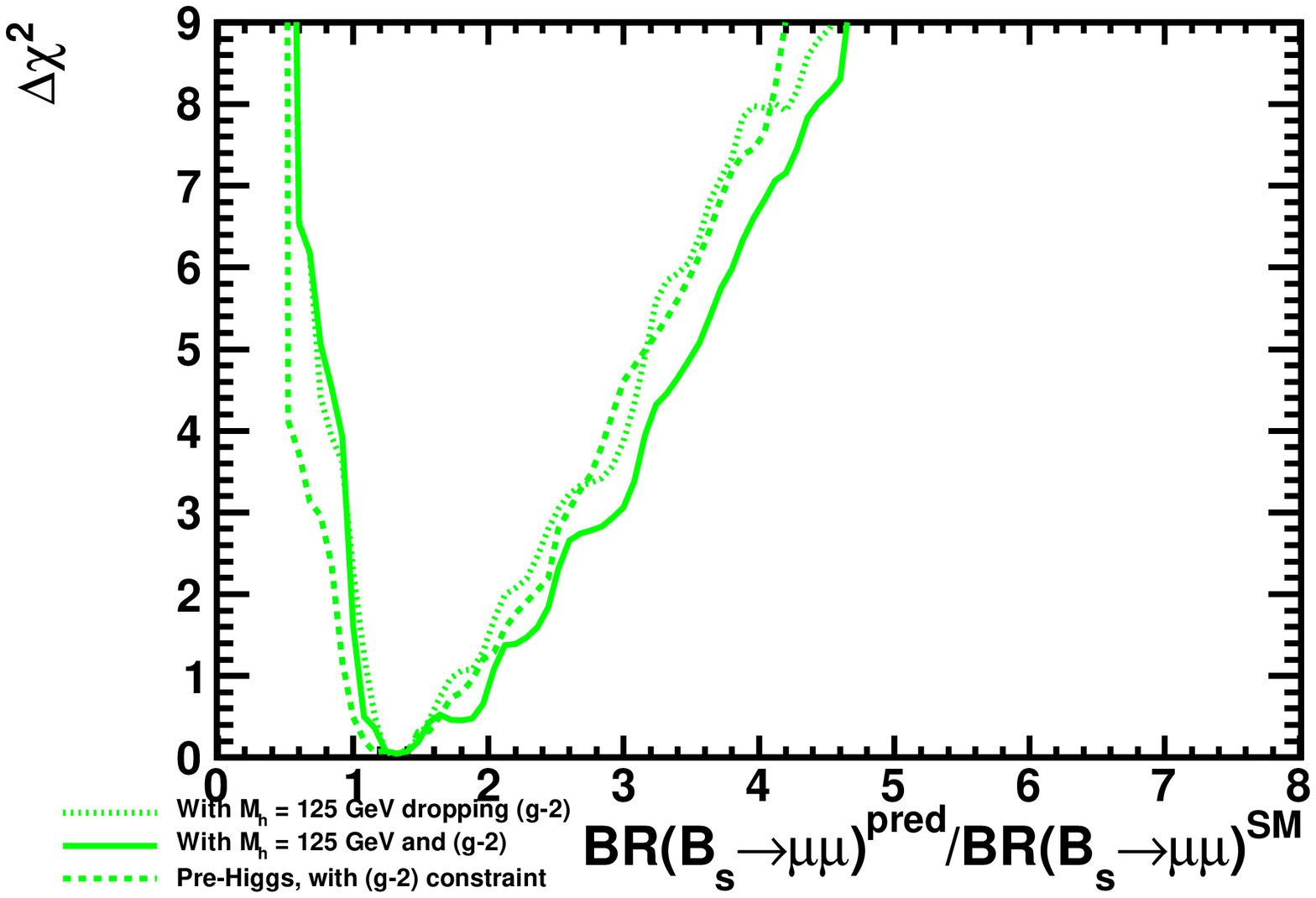}}
\resizebox{8.2cm}{!}{\includegraphics{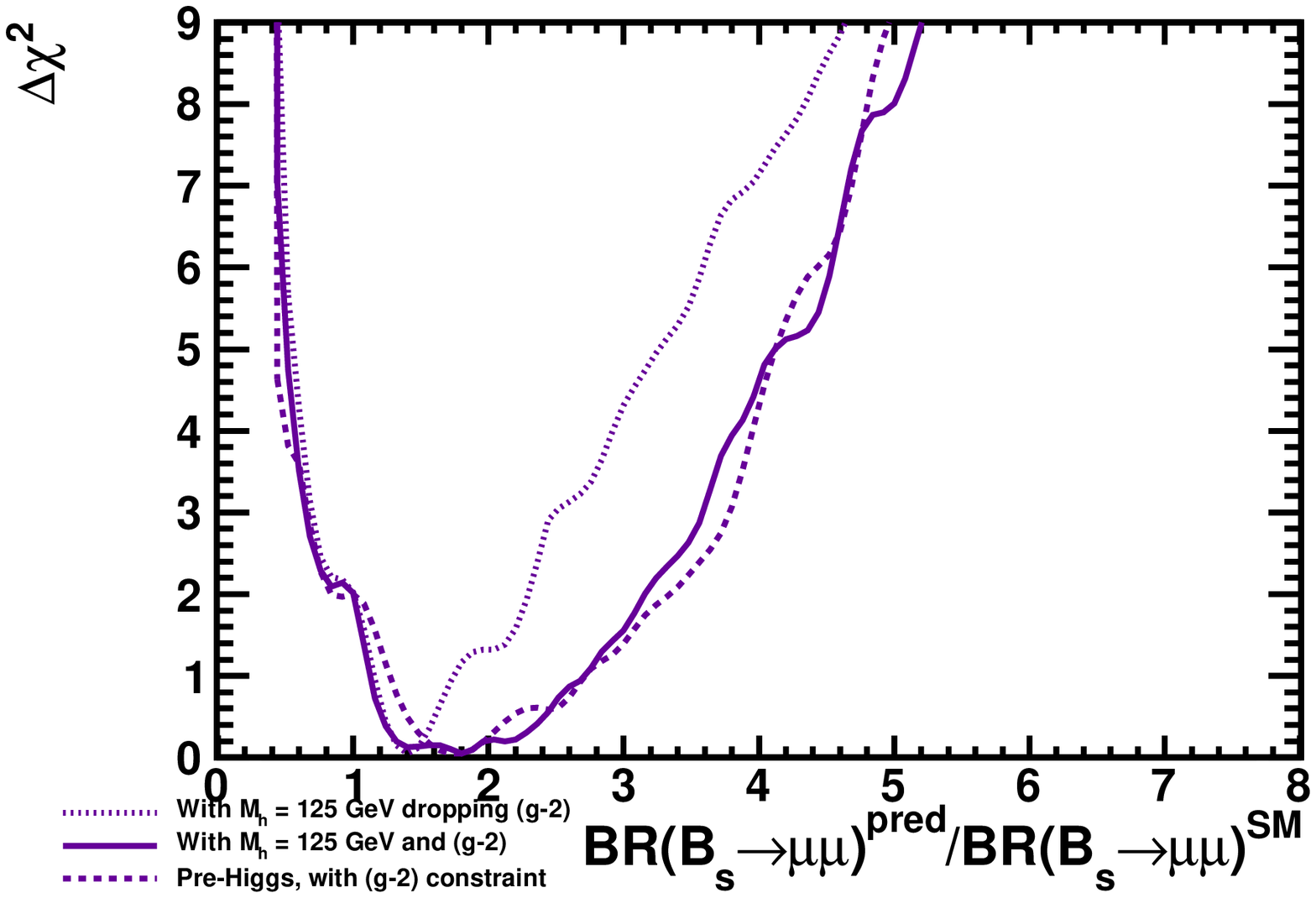}}
\vspace{-1cm}
\caption{\it The one-dimensional $\Delta \chi^2$ functions for \bmm\ in
the CMSSM (left) and the NUHM1 (right), for $\Mh \simeq 125 \gev$. The
notations and significations of 
the lines are the same as in Fig.~\protect\ref{fig:mgl}.
}
\label{fig:bmm}
\end{figure*}

We now turn to the predictions of our fits for \bmm, shown in
Fig.~\ref{fig:bmm}. This observable is not very sensitive directly to $\Mh$,
and the indirect sensitivity via $m_{1/2}$ is not very strong, though
smaller values of $m_{1/2}$ do lead to larger values of \bmm, in general.
As seen in Fig.~\ref{fig:bmm}, imposing the putative LHC $\Mh$ constraint
indeed has little effect on \bmm. We recall that the best-fit values in
the CMSSM and NUHM1 are both slightly larger than in the SM,
and enhancements of up to \order{30-40\%} with respect to the SM
prediction could be detected at the LHC at the $3\,\sigma$~level.

\begin{figure*}[htb!]
\resizebox{8.6cm}{!}{\includegraphics{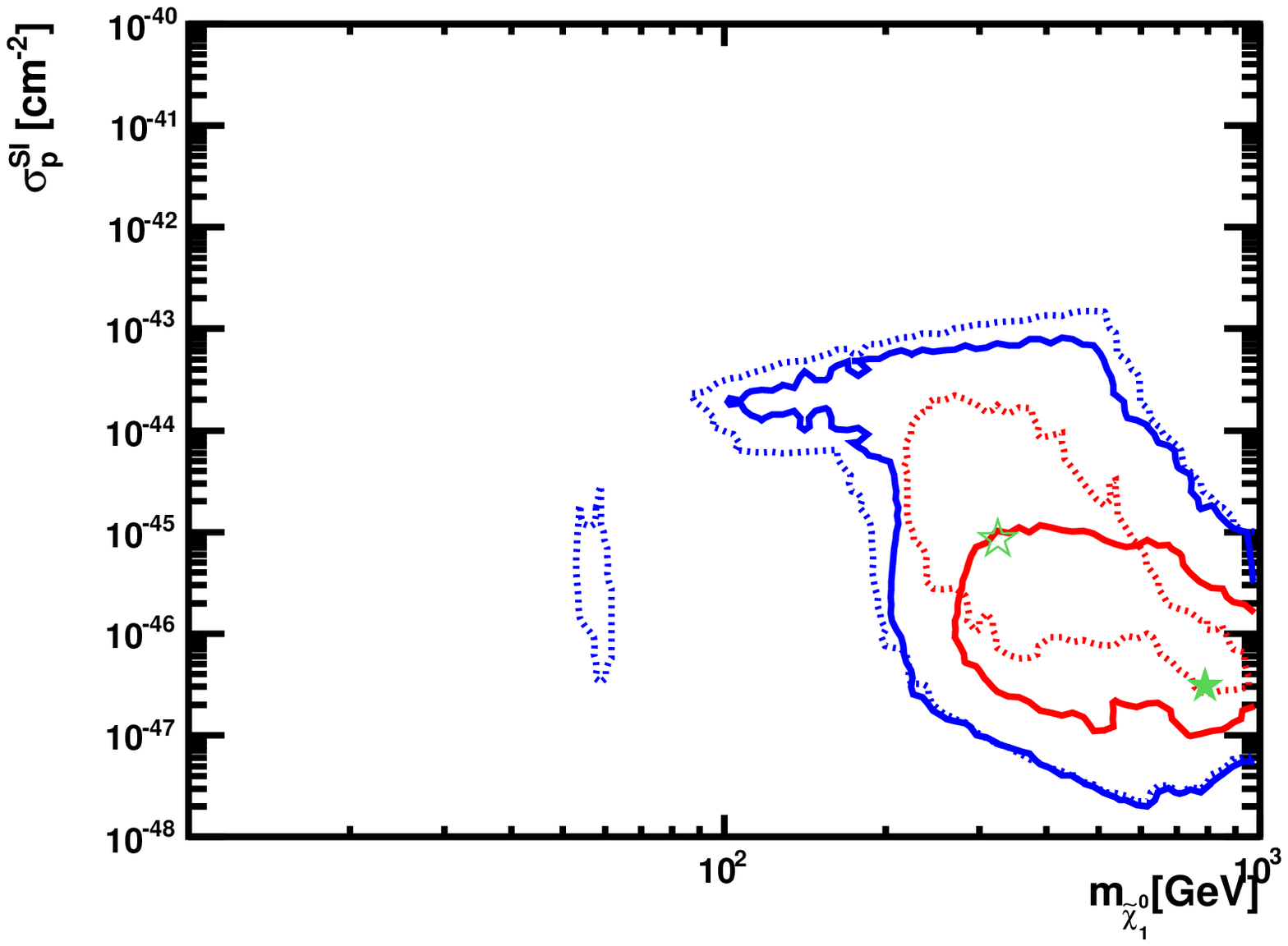}}
\resizebox{8.6cm}{!}{\includegraphics{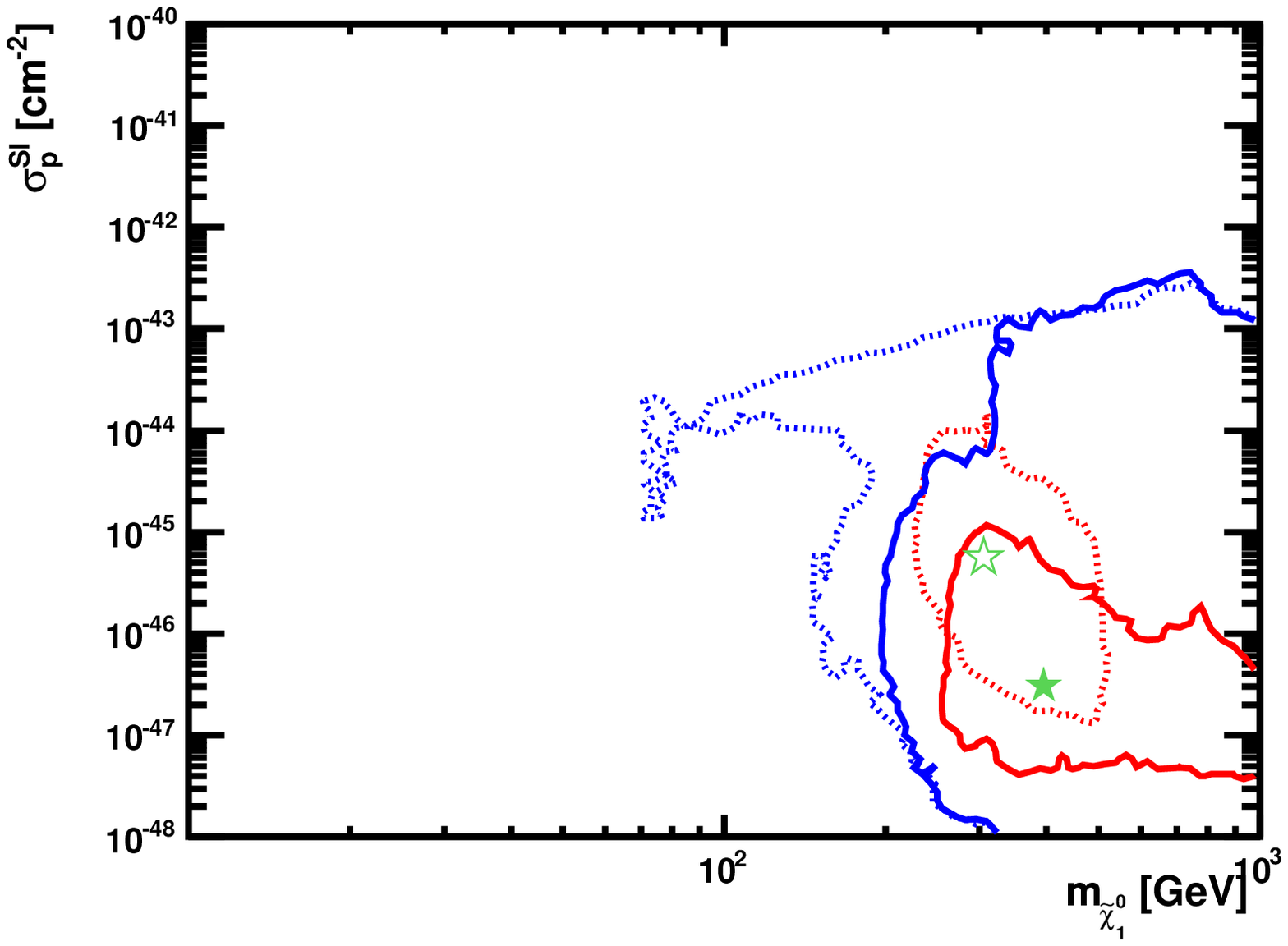}}
\vspace{-1cm}
\caption{\it The $(\mneu{1}, \ssi)$ planes in the CMSSM (left) and the
  NUHM1 (right), for $\Mh \simeq 125 \gev$. 
  The notations and significations of the contours are
  the same as in Fig.~\protect\ref{fig:6895}. 
}
\label{fig:mneussi}
\end{figure*}

Finally, in Fig.~\ref{fig:mneussi} we show results for the
preferred regions in the $(\mneu{1}, \ssi)$ plane. 
As seen in Fig.~\ref{fig:mneussi}, the fact that larger values of $m_{1/2}$
and hence $\mneu{1}$ are favoured by the larger values of $\Mh$ implies that
at the 68\% CL the preferred range of \ssi\ is significantly lower when
$\Mh \simeq 125 \gev$, when compared to our previous best fit
with $\Mh = 119 \gev$,  rendering direct detection of dark matter
significantly more difficult.
Again, this effect on $\mneu{1}$ is more pronounced in the CMSSM, whereas in the NUHM1 
the value of $\mneu{1}$ for the best-fit point changes only
  slightly.

\subsubsection*{Results dropping the \boldmath{\gmt} constraint}

\noindent\\
We have restricted our attention so far to $\Mh \simeq 125 \gev$
assuming the \gmt\ constraint. However, this value of $\Mh$
corresponds approximately to our best-fit points in \cite{mc7} when 
the \gmt\ constraint is dropped~%
\footnote{We recall that it was shown in~\cite{mc7} that the
CMSSM/NUHM1 interpretation of \gmt\ is in
some tension with the LHC constraints on events with $\ETslash$.}. 
Accordingly, we now consider an the same measurement as
given in Eq.~(\ref{Mh125}), but with \gmt\ dropped from the fit~%
\footnote{There are small differences between the pre-Higgs 68 and 95\%
CL contours presented here and the corresponding contours in~\cite{mc7}, 
which provide a measure of the uncertainties in the interpretation of
the MCMC data generated for our analysis.}. 
In the following plots we show results for fits omitting \gmt, pre-Higgs (dotted)
and post-Higgs (solid).

\begin{figure*}[htb!]
\resizebox{8.6cm}{!}{\includegraphics{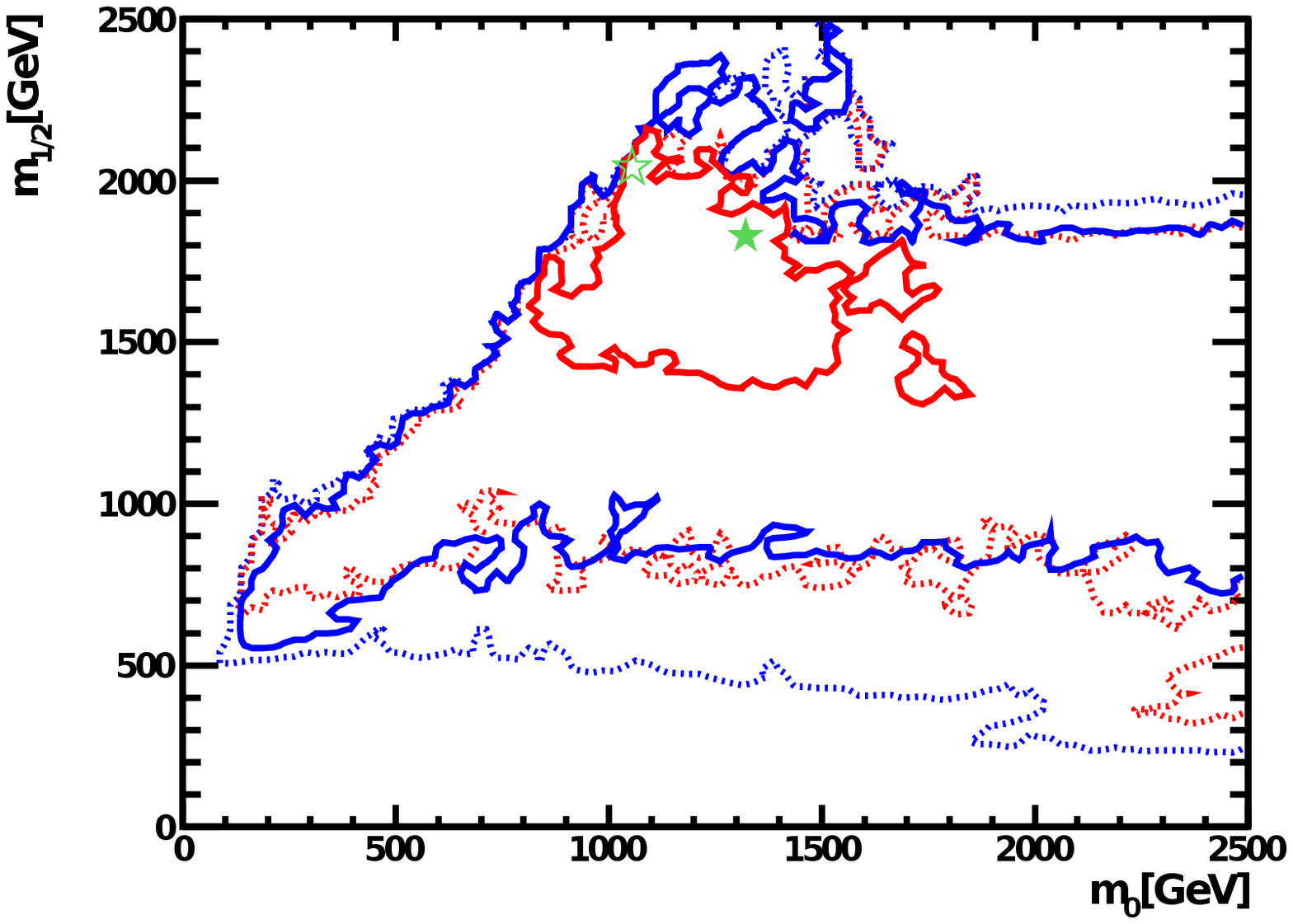}}
\resizebox{8.6cm}{!}{\includegraphics{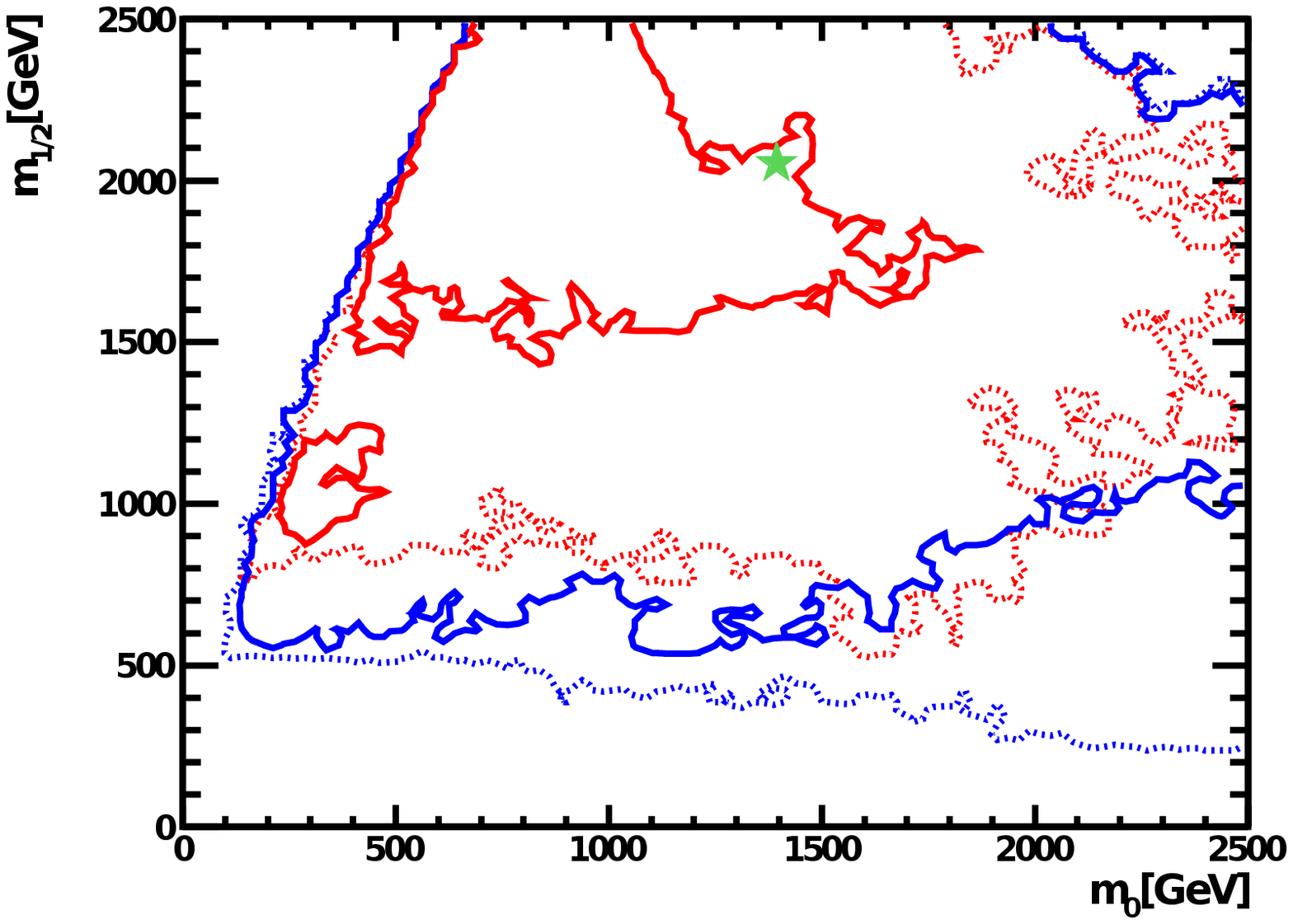}}
\caption{\it The $(m_0, m_{1/2})$ planes in the CMSSM (left) and the
  NUHM1 (right), for $\Mh \simeq 125 \gev$, but dropping the \gmt\
  constraint.   Dotted lines show the contours
  found previously in~\protect\cite{mc7} dropping the \gmt\ but without this $\Mh$ constraint.
  Here the open green stars denote the pre-Higgs best-fit
  points~\protect\cite{mc7} (also dropping \gmt), whereas the solid green
  stars indicate the new best-fit points. These best-fit points are essentially coincident in the NUHM1 case.
}
\label{fig:6895-nogmt}
\end{figure*}

We see in Fig.~\ref{fig:6895-nogmt} that the regions of the 
$(m_0, m_{1/2})$ planes in 
the CMSSM and NUHM1 that are favoured at the 68\% CL are concentrated at large
values if the \gmt\ constraint is dropped. This reflects the relative
harmony between the LHC $\ETslash$ constraints and the hypothetical 
$\Mh \simeq 125 \gev$ measurement 
if \gmt\ is omitted. The inclusion of Eq.~(\ref{Mh125}) substantially
  sharpens the prediction at the 68\%~CL, whereas it is less pronounced for
  the 95\%~CL contours.

\begin{figure*}[htb!]
\resizebox{8.6cm}{!}{\includegraphics{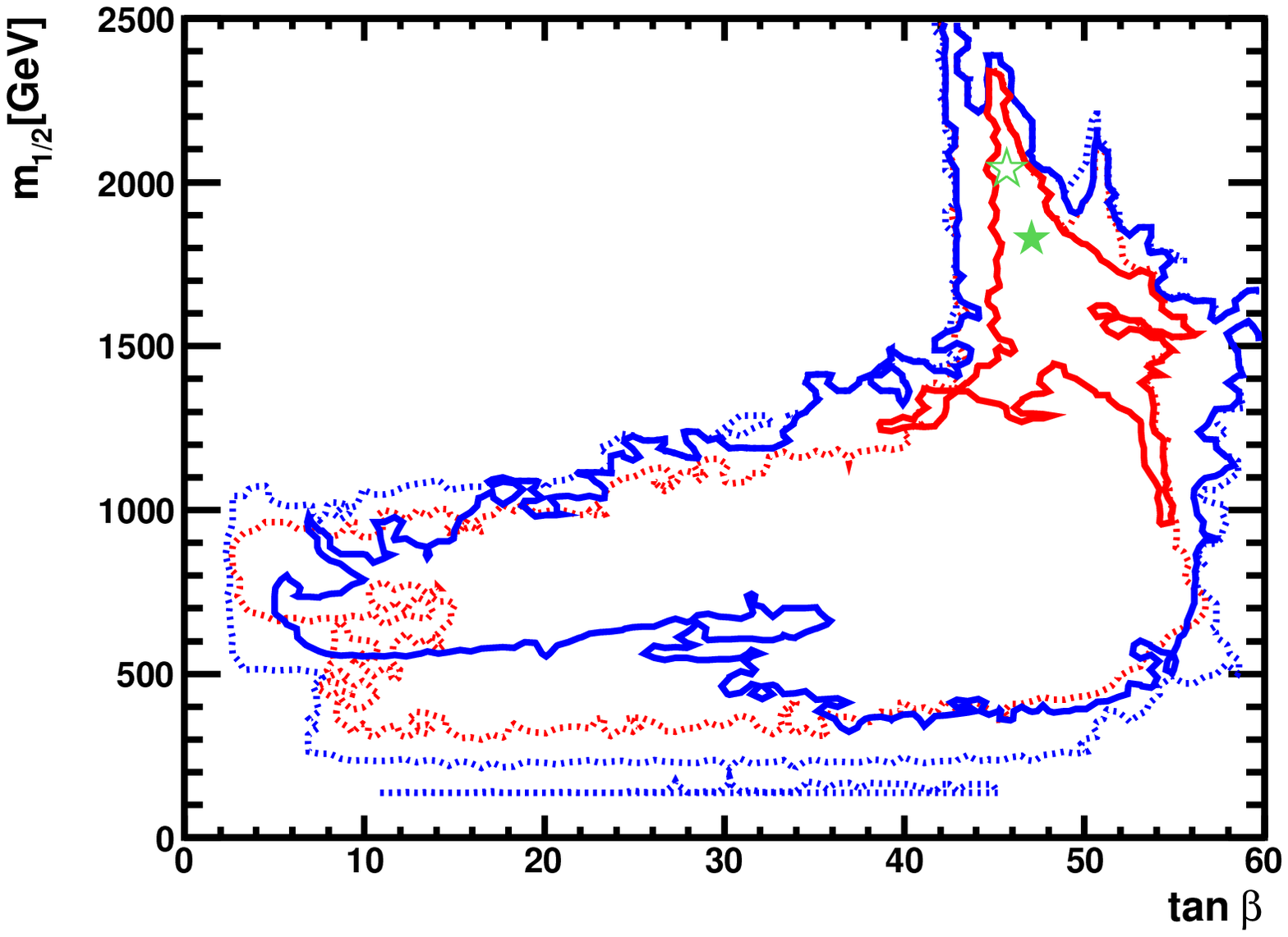}}
\resizebox{8.6cm}{!}{\includegraphics{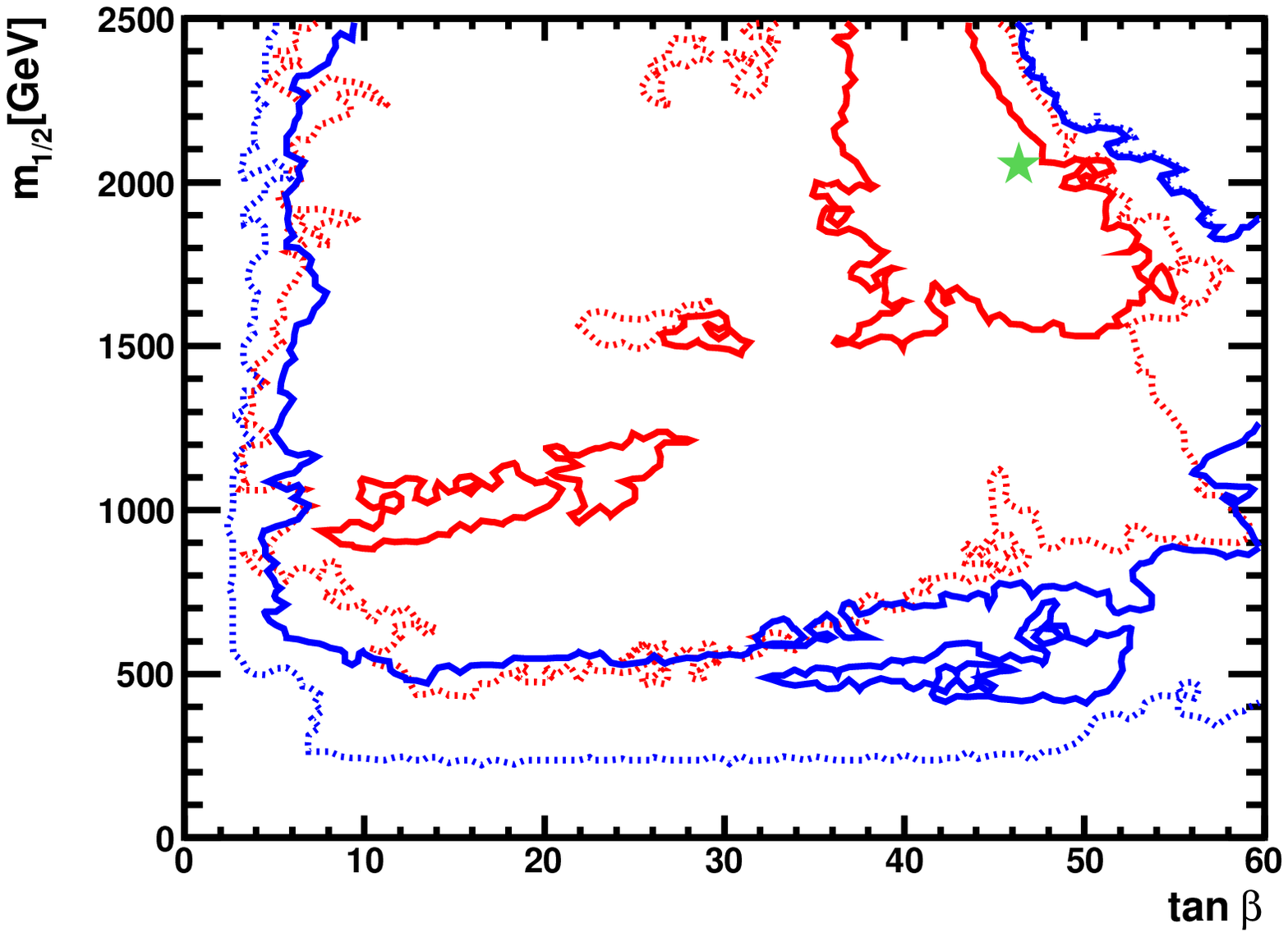}}
\vspace{-1cm}
\caption{\it The $(m_{1/2}, \tb)$ planes in the CMSSM (left) and the
  NUHM1 (right), for $\Mh \simeq 125 \gev$, but dropping the \gmt\
  constraint. The notations and significations of the contours are 
  the same as in Fig.~\protect\ref{fig:6895-nogmt}.
}
\label{fig:tanbm12-nogmt}
\end{figure*}

\begin{figure*}[htb!]
\resizebox{8.6cm}{!}{\includegraphics{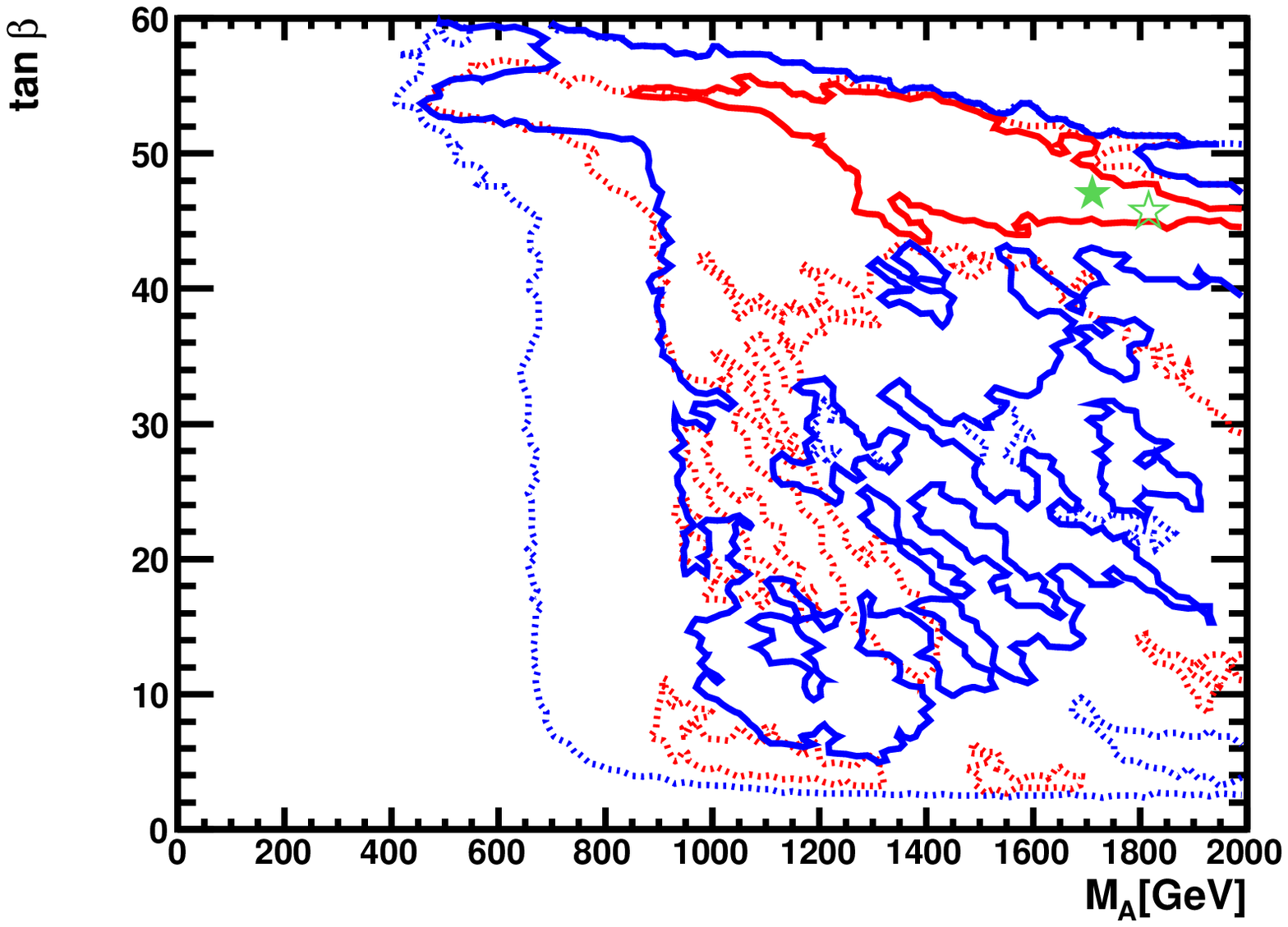}}
\resizebox{8.6cm}{!}{\includegraphics{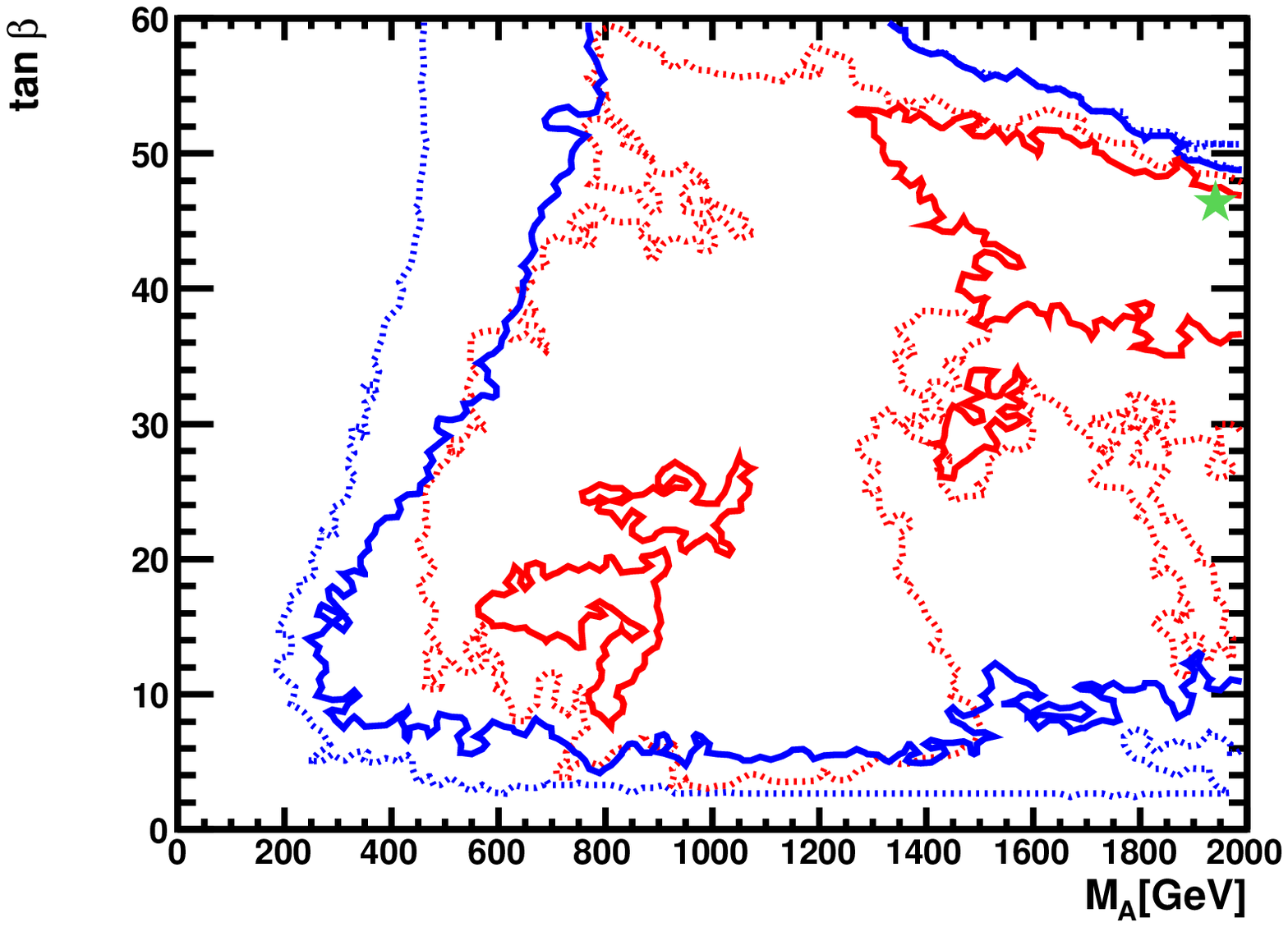}}
\vspace{-1cm}
\caption{\it The $(\MA, \tb)$ planes in the CMSSM (left) and the
  NUHM1 (right), for $\Mh \simeq 125 \gev$, but dropping the \gmt\ constraint. The notations and significations of the contours are
  the same as in Fig.~\protect\ref{fig:6895-nogmt}.
}
\label{fig:MAtb-nogmt}
\end{figure*}

As we see in Fig.~\ref{fig:tanbm12-nogmt}, the concentration at relatively
large $m_{1/2}$ is reflected in a correlated preference for large values
of $\tb$. Furthermore, as seen in Fig.~\ref{fig:MAtb-nogmt}, the
corresponding preferred range of 
$\MA$ is also concentrated at relatively large masses.
Again the inclusion of Eq.~(\ref{Mh125}) considerably sharpens the
  preferred parameter ranges.

Looking back now at the one-dimensional $\Delta \chi^2$ functions for the fits
without \gmt\ that are shown as dotted lines in Figs.~\ref{fig:mgl} and
\ref{fig:mstau}, we see that the preference for large values of 
$(m_0, m_{1/2})$ carries over into relatively large values 
of $\mgl, \msqr$ and $\mstaue$. In particular, the \gmt-less scenarios
offer somewhat gloomy prospects for sparticle detection at the LHC. On
the other hand, as seen in Figs.~\ref{fig:bmm}, there is little change
in the prediction for \bmm\ if \gmt\ is omitted.

\begin{figure*}[htb!]
\resizebox{8.6cm}{!}{\includegraphics{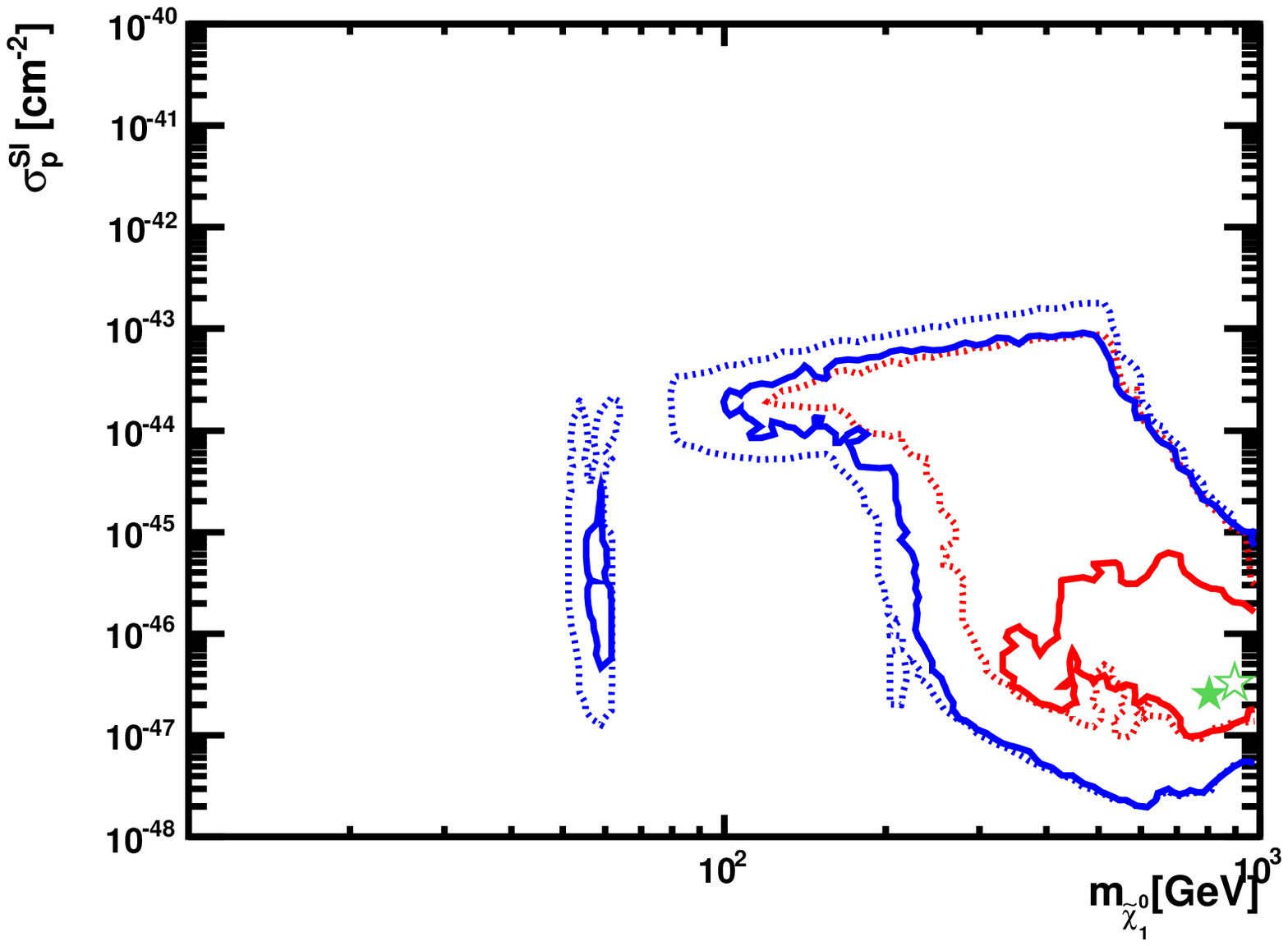}}
\resizebox{8.6cm}{!}{\includegraphics{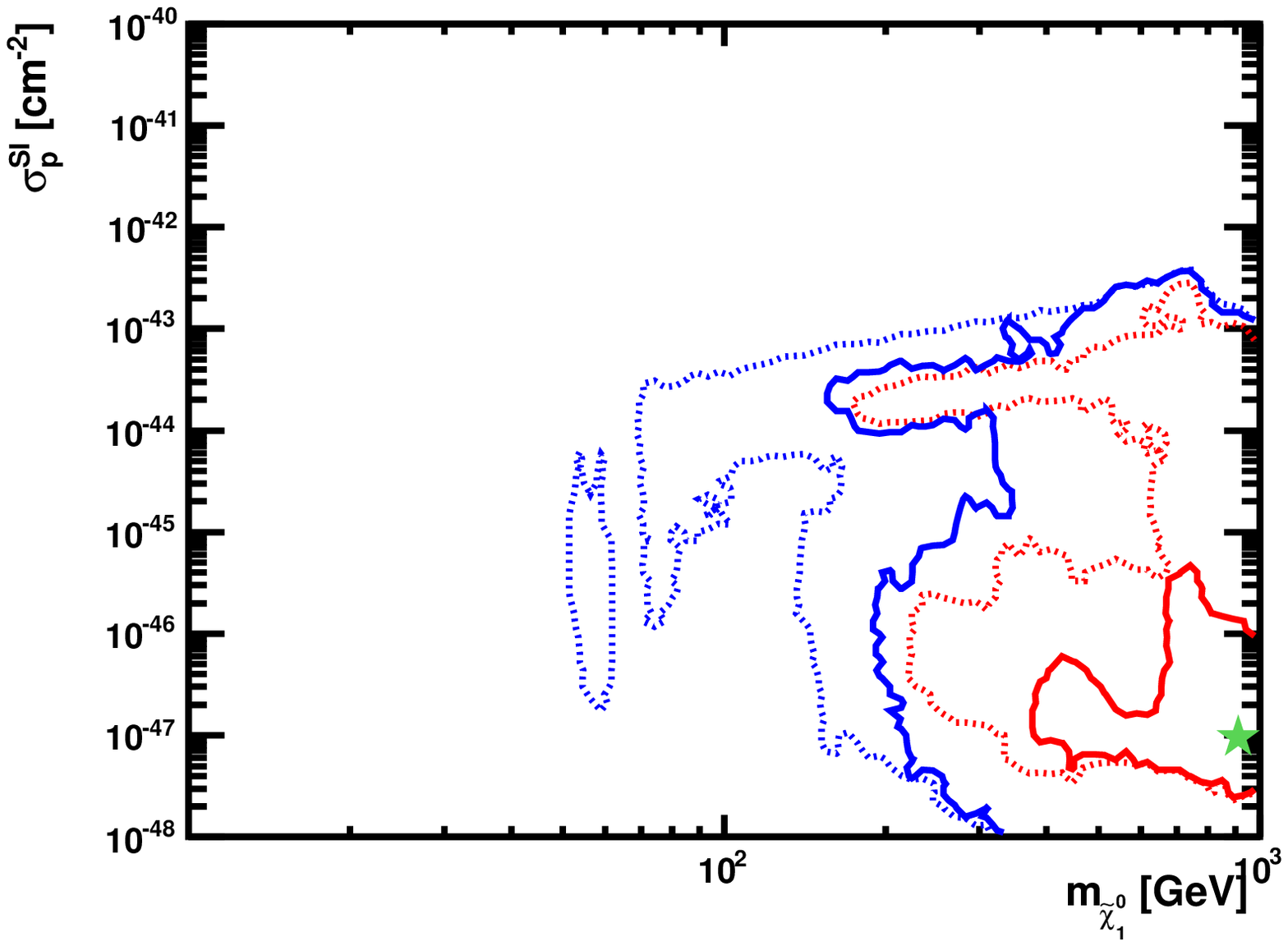}}
\vspace{-1cm}
\caption{\it The $(\mneu{1}, \ssi)$ planes in the CMSSM (left) and the
  NUHM1 (right), for $\Mh \simeq 125 \gev$, but dropping the \gmt\
  constraint. 
  The notations and significations of the contours are
  the same as in Fig.~\protect\ref{fig:6895-nogmt}. 
}
\label{fig:mneussi-nogmt}
\end{figure*}

Turning finally to the predictions for \ssi\ if \gmt\ is omitted, shown
in Fig.~\ref{fig:mneussi-nogmt}, 
we see that the relatively large values of $m_{1/2}$ seen in
Fig.~\ref{fig:6895-nogmt} are reflected  in relatively large values of
$\mneu{1}$, which correspond in turn to relatively low values of \ssi.
The inclusion of Eq.~(\ref{Mh125}) again strongly reduces the preferred
parameter ranges.

An alternative interpretation of a Higgs signal around $\Mh \simeq 125 \gev$
would be that while the MSSM might still be realized, it is not the
CMSSM nor the NUHM1 that describes Nature correctly, but another
version of the MSSM. In this case, the prospects for sparticle detection
at the LHC 
and dark matter detection might well be more cheerful than in the \gmt-less
CMSSM and NUHM1 scenarios discussed here.
However, the exploration of such possible alternative models lies beyond
the scope of our analysis. 


\subsubsection*{What if \boldmath{$\Mh = 119 \gev$}?}

\noindent\\
We have restricted our attention so far to $\Mh \simeq 125 \gev$, which 
corresponds to the excess seen in both CMS and ATLAS. 
We now consider an alternative potential
LHC measurement $\Mh = 119 \pm 1 \gev$, which corresponds to the CMS
$ZZ^*$ signal and our earlier predictions including the \gmt\ constraint, again
allowing for a theoretical error 
$\pm 1.5 \gev$ in the calculation of $\Mh$ for any given set of CMSSM or NUHM1
parameters. 

\begin{figure*}[htb!]
\resizebox{8.6cm}{!}{\includegraphics{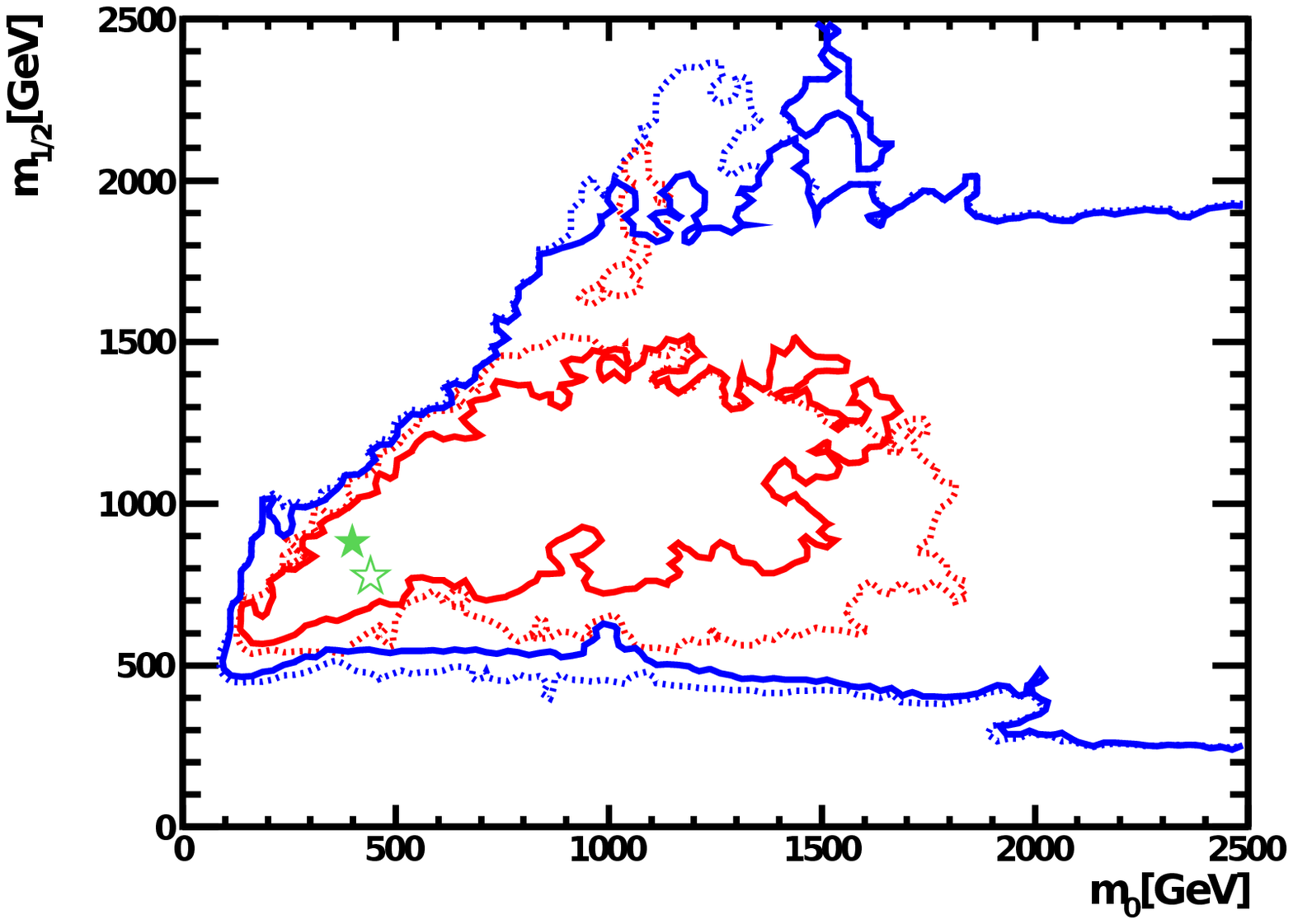}}
\resizebox{8.6cm}{!}{\includegraphics{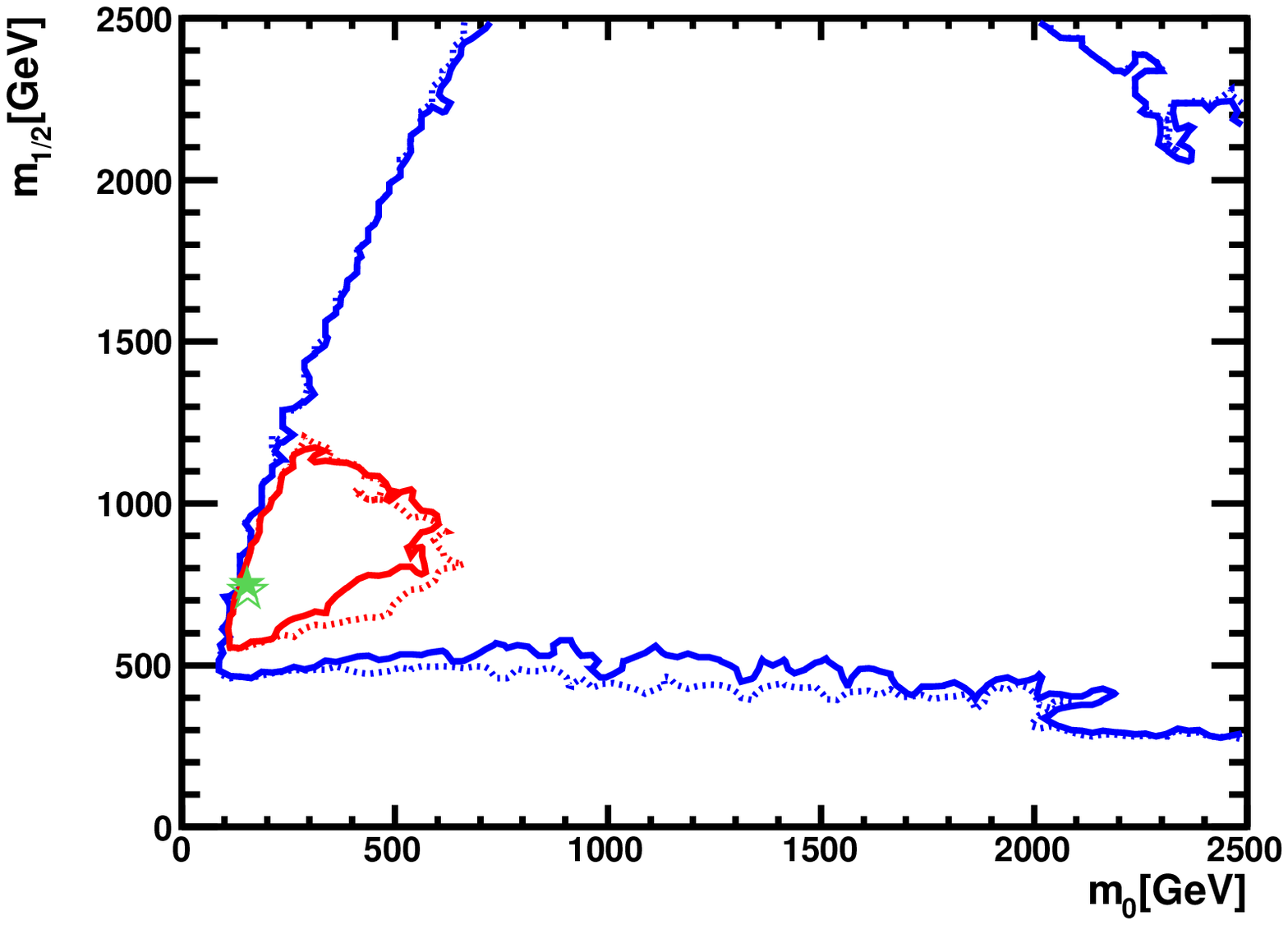}}
\caption{\it The $(m_0, m_{1/2})$ planes in the CMSSM (left) and the
  NUHM1 (right) assuming a hypothetical measurement of $\Mh = 119 \gev$.   
  The notations and significations of the contours are
  the same as in Fig.~\protect\ref{fig:6895}. 
}
\label{fig:6895119}
\end{figure*}

The $(m_0, m_{1/2})$ planes
shown in Fig.~\ref{fig:6895119} for the CMSSM (left) and NUHM1 (right),
the preferred regions are shown at the 68\%~CL (red) and
95\%~CL, with the solid (dotted) lines include (omit) the assumed LHC
Higgs constraint.
Since this assumed LHC
value of $\Mh$ coincides with the previous best-fit values in both
the CMSSM and NUHM1, 
the best-fit points in these models
(indicated by the green stars in Fig.~\ref{fig:6895119}) are nearly
unaffected by the imposition of the putative LHC constraint.~%
The effect of the hypothetical measurement restricting the range 
in $m_{1/2}$ is indeed seen in both panels of Fig.~\ref{fig:6895119}, 
though
for the 68\% CL contour (shown in red) it is much more pronounced for
the CMSSM than for the NUHM1, whereas for the 95\% CL contour (shown in
blue) it is more significant for the NUHM1. This reflects the fact in
the NUHM1 the global $\Delta \chi^2$ function found 
in~\cite{mc7} rose quite steeply in the neighbourhood of the best-fit point,
resulting in a relatively tight 68\% CL contour, whereas the rise of $\chi^2$
away from the best-fit point in the CMSSM was more gradual. This led
previously to a larger 68\% CL contour and a broader range of $\Mh$ at
the 68\% CL, which is now more affected by an assumed LHC $\Mh$ constraint. On
the other hand, the 95\% CL contour in the NUHM1 extended previously to larger
values of $m_{1/2}$ than in the CMSSM, and these values are particularly
susceptible to the LHC $\Mh$ constraint. 

\begin{figure*}[htb!]
\resizebox{8.6cm}{!}{\includegraphics{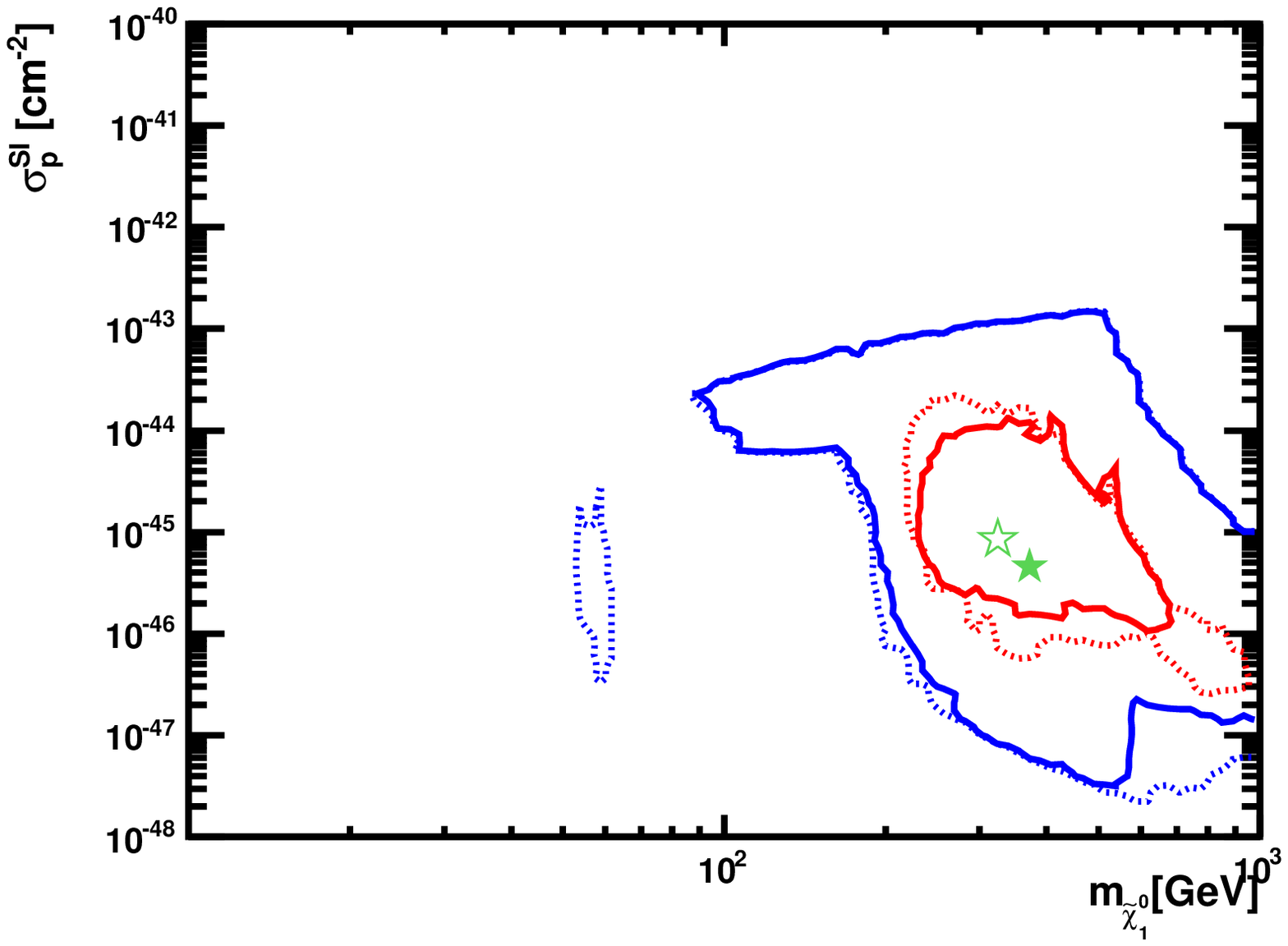}}
\resizebox{8.6cm}{!}{\includegraphics{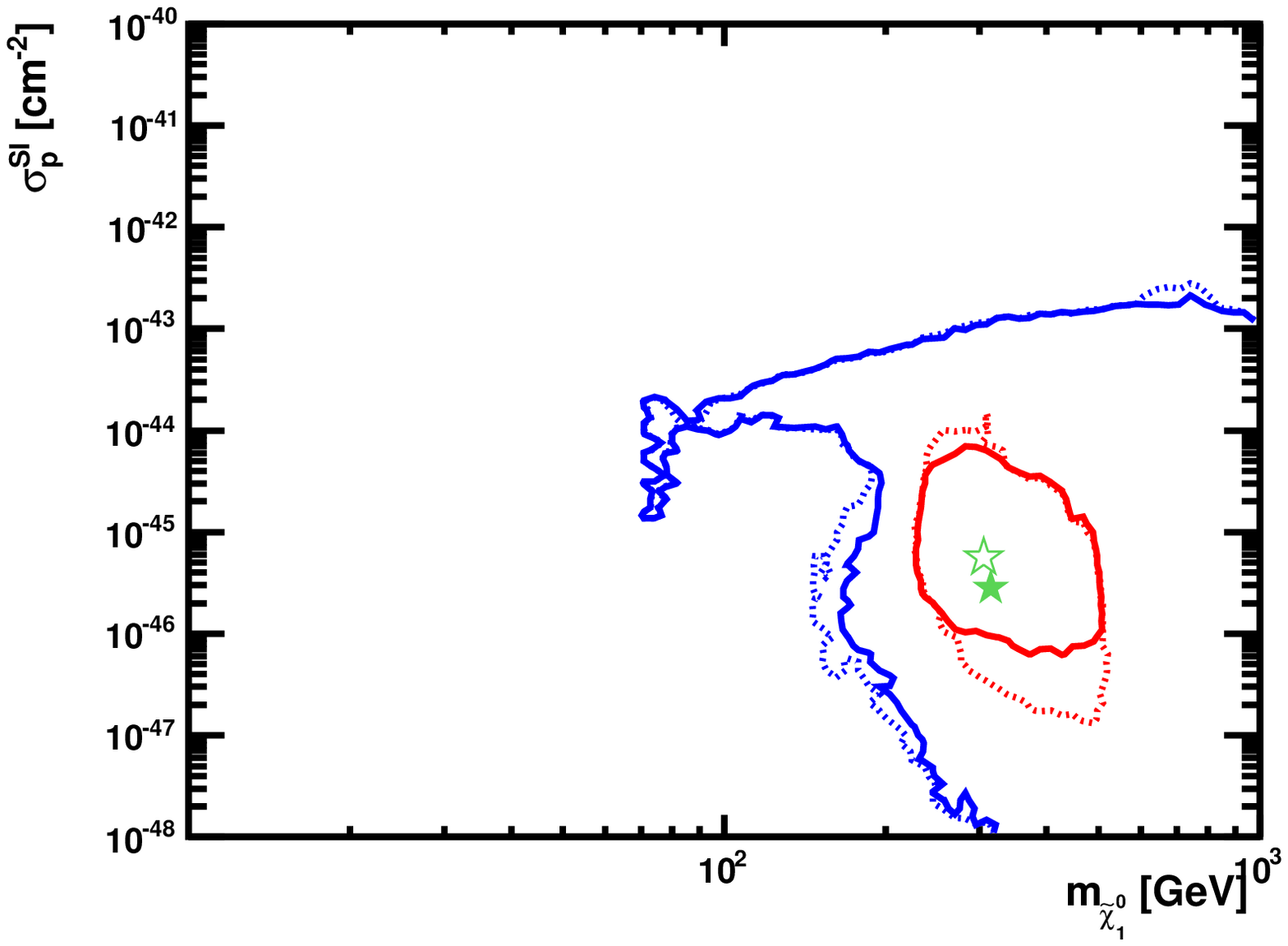}}
\vspace{-1cm}
\caption{\it The $(\mneu{1}, \ssi)$ planes in the CMSSM (left) and the
  NUHM1 (right), for $\Mh \simeq 119 \gev$. 
  The notations and significations of the contours are
  the same as in Fig.~\protect\ref{fig:6895119}. 
}
\label{fig:mneussi119}
\end{figure*}

Although we add another constraint (as discussed above),  the total $\chi^2$
at the best-fit points do not change~\footnote{They would change only slightly
if the Higgs mass were assumed to differ by $\lsim 1 \gev$ from that obtained at
the best-fit point.}. For this reason, the $p$-values for the CMSSM and NUHM1
would increase for a hypothetical measurement $\Mh \simeq 119 \gev$, 
corresponding formally to better overall fits to the larger data set, as seen in Table~\ref{tab:p-comp}.

As one might expect, such an LHC $\Mh$ constraint would reduce
considerably the 68\% CL range of $\tb$ in the CMSSM. This is because,
for $m_{1/2}$ close to the best-fit value, $\sim 700$ to $800 \gev$,
fixing the Higgs mass at $119 \gev$ disfavours low values of $\tb$,
which yield low values of $\Mh$. This effect is not important in the
NUHM1, where the range of $\tb$ was already smaller before imposing the
$\Mh$ constraint. We also note that $\MA$
is restricted to somewhat smaller values when
the hypothetical LHC constraint on $\Mh$ is included.
Furthermore, as expected,
the values of $\mgl$ at the minima of the $\chi^2$ functions are
not affected and there is little change in $\chi^2$ for $\mgl$
between 2 and 3~TeV. (The corresponding plots are not shown.)
However,  there are significant effects at both lower
and higher values of $\mgl$. In particular, large values of $\mgl \gsim 3 \tev$ are disfavoured.
The prospects for discovering gluinos
at the LHC in the near future would remain uncertain in both the CMSSM
and NUHM1.
An LHC measurement of $\Mh \simeq 119 \gev$
would disfavour large squark masses, but the 95\% CL range would still
extend to $\msqr \sim 4 \tev$ in the CMSSM and $\sim 2 \tev$ in the NUHM1.
The preferred value of $\mstaue \sim 300 \gev$ in both the
CMSSM and NUHM1 both with and without the hypothetical LHC $\Mh$
measurement, with large masses again becoming somewhat more disfavoured.

Finally, in both the CMSSM
and the NUHM1 there is little impact on the 95\% CL regions  
nor on the 68\% CL region in the NUHM1 in the ($m_{\mneu{1}},\ssi)$ plane. 
The only substantial change, as can be seen in
  Fig.~\ref{fig:mneussi119}, appears in the 68\% CL region of the CMSSM,
where now values of $\mneu{1} \gsim 700 \gev$ and
$\ssi \lsim 10^{-46} {\rm cm}^{-2}$ are disfavoured after the inclusion of a
Higgs-boson mass measurement at $119 \gev$.


\section{Summary and Conclusions}

The ATLAS and CMS searches for the Higgs boson have already excluded a very
large range of masses, with the only remaining windows for a SM-like
Higgs boson being in the ranges $\Mh \in (115.5, 127) \gev$ or $> 600
\gev$~\cite{ATLAS+CMS,Dec13}. The latter range is disfavoured by precision
electroweak data, so attention naturally focuses on the low-mass
range. It may or not be a coincidence that this range includes the range 
$\Mh \lsim 130 \gev$ accessible in simple supersymmetric models
such as the CMSSM and NUHM1. Within this range, our previous global fits
of these models including \gmt\ predicted $\Mh \sim 119 \gev$ if the \gmt\
constraint was included in the fit, and $\Mh \sim 126 \gev$ if \gmt\
was omitted~\cite{mc7}. The 
latest ATLAS and CMS results
display an interesting fluctuation at $\Mh \sim 125 \gev$ (i.e.\
close to the latter result from \cite{mc7}) and we
have combined a hypothetical measurement of 
$\Mh = 125 \gev$ with the global likelihood functions obtained in our
previous fits~\cite{mc7}. 

As we have shown in this paper, this combination refines our previous
predictions for the CMSSM and NUHM1 model parameters within
global fits incorporating \gmt. In particular, the combination prefers 
a range of larger values of $m_{1/2}$, resulting in larger values of 
$\mgl$ and other sparticle masses being preferred, restricting the prospects for
discovering supersymmetry at the LHC within these models. 
The predictions for \ssi\ are pushed to higher masses and lower 
cross sections, particularly in the CMSSM. There are also
smaller changes in the predictions for other observables such as
\bmm\ .

We have also shown the analogous CMSSM and NUHM1 fit results for a hypothetical
measurement of $\Mh \simeq 125 \gev$ if the \gmt\ constraint is omitted.
In this case we find a stronger preference for larger values of $(m_0, m_{1/2})$, 
and correspondingly larger values of $\tb$ and $\MA$, as well as larger
values of $\mgl, \msqr$,
potentially lying beyond the reach of the LHC. We have also
commented on the potential implications of a hypothetical Higgs
discovery at $\Mh \simeq 119 \gev$.

Time will soon tell where the LHC experiments are indeed discovering the
Higgs boson. However, we have shown that $\Mh = 125 \gev$ is a
possibility within the CMSSM and NUHM1, although it lies at the upper
range of what is possible within the CMSSM or NUHM1, and might suggest
reduced prospects for discovering these particular models of
supersymmetry at the LHC. Alternatively, it could well be that one
should look beyond the frameworks of the models discussed here. 


\subsubsection*{Note Added}

After acceptance of this paper for publication, we became aware of issues
in the implementation of the {\tt FeynHiggs} code and in the cold dark matter
density calculation, which required extra sampling and reprocessing of the NUHM1
parameter space. We are grateful to Nazila~Mahmoudi and Azar Mustafayev
for discussions on dark matter density calculations.

\subsubsection*{Acknowledgements}

The work of O.B., K.J.D., J.E., J.M. and K.A.O. is supported in part by
the London Centre for Terauniverse Studies (LCTS), using funding from
the European Research Council 
via the Advanced Investigator Grant 267352. 
The work of S.H. is supported 
in part by CICYT (grant FPA 2010--22163-C02-01) and by the
Spanish MICINN's Consolider-Ingenio 2010 Program under grant MultiDark
CSD2009-00064. The work of K.A.O. is supported in part by DOE grant
DE-FG02-94ER-40823 at the University of Minnesota.



\begin{thebibliography}{99}


 \bibitem{Ellis:2001qv}
  J.~R.~Ellis, S.~Heinemeyer, K.~A.~Olive and G.~Weiglein,
  Phys.\ Lett.\  B {\bf 515} (2001) 348
  [arXiv:hep-ph/0105061].

\bibitem{Ambrosanio:2001xb}
  S.~Ambrosanio, A.~Dedes, S.~Heinemeyer, S.~Su and G.~Weiglein,
  Nucl.\ Phys.\  B {\bf 624} (2002) 3
  [arXiv:hep-ph/0106255].

\bibitem{mc7}
 O.~Buchmueller, {\it et al.},
  Eur.\ Phys.\ J.\ C {\bf 72}, 1878 (2012)
  [arXiv:1110.3568 [hep-ph]].

\bibitem{mh}
J.~R.~Ellis, G.~Ridolfi and F.~Zwirner,
  Phys.\ Lett.\ B {\bf 257} (1991) 83;
  Phys.\ Lett.\ B {\bf 262} (1991) 477;
  Yasuhiro Okada, Masahiro Yamaguchi and Tsutomu Yanagida,
{Phys. Lett.} B262, 54, 1991;
{Prog. Theor. Phys.} 85 (1991) 1;
  A.~Yamada,
  Phys.\ Lett.\ B {\bf 263} (1991) 233;
  Howard~E. Haber and Ralf Hempfling,
{Phys. Rev. Lett.} 66 (1991) 1815;
M.~Drees and M.~M.~Nojiri,
  Phys.\ Rev.\ D {\bf 45} (1992) 2482;
  P.~H.~Chankowski, S.~Pokorski and J.~Rosiek,
  Phys.\ Lett.\ B {\bf 274} (1992) 191;
  Phys.\ Lett.\ B {\bf 286} (1992) 307.

  \bibitem{FeynHiggs}
 G.~Degrassi, S.~Heinemeyer, W.~Hollik, P.~Slavich and G.~Weiglein,
  Eur.\ Phys.\ J.\ C {\bf 28} (2003) 133
  [arXiv:hep-ph/0212020];
   S.~Heinemeyer, W.~Hollik and G.~Weiglein,
  Eur.\ Phys.\ J.\ C {\bf 9} (1999) 343
  [arXiv:hep-ph/9812472];
  S.~Heinemeyer, W.~Hollik and G.~Weiglein,
  Comput.\ Phys.\ Commun.\  {\bf 124} (2000) 76
  [arXiv:hep-ph/9812320];
   M.~Frank {\it et al.}, 
  JHEP {\bf 0702} (2007) 047
  [arXiv:hep-ph/0611326];
  See {\tt http://www.feynhiggs.de}~.

\bibitem{asbstev} 
M.~Carena, P.~Draper, S.~Heinemeyer, T.~Liu, C.~E.~M.~Wagner
  and G.~Weiglein, 
  Phys.\ Rev.\ D {\bf 83} (2011) 055007
  [arXiv:1011.5304 [hep-ph]].


\bibitem{mc1}
O.~Buchmueller {\it et al.},
  Phys.\ Lett.\  B {\bf 657} (2007) 87
  [arXiv:0707.3447 [hep-ph]].

\bibitem{mc2}
  O.~Buchmueller {\it et al.},
  JHEP {\bf 0809} (2008) 117
  [arXiv:0808.4128 [hep-ph]].

\bibitem{mc3}
  O.~Buchmueller {\it et al.},
  Eur.\ Phys.\ J.\  C {\bf 64} (2009) 391
  [arXiv:0907.5568 [hep-ph]].

\bibitem{mc35}
  O.~Buchmueller {\it et al.},
  Phys.\ Rev.\  D {\bf 81} (2010) 035009
  [arXiv:0912.1036 [hep-ph]].

\bibitem{mc4}
  O.~Buchmueller {\it et al.},
  Eur.\ Phys.\ J.\  C {\bf 71} (2011) 1583
  [arXiv:1011.6118 [hep-ph]].

\bibitem{mc5}
  O.~Buchmueller {\it et al.},
  Eur.\ Phys.\ J.\  C {\bf 71} (2011) 1634
  [arXiv:1102.4585 [hep-ph]].

\bibitem{mc6}
O.~Buchmueller {\it et al.},
  Eur.\ Phys.\ J.\  C {\bf 71} (2011) 1722
  [arXiv:1106.2529 [hep-ph]].

\bibitem{mcweb}
For more information and updates, please see {\tt http://cern.ch/mastercode/}.

\bibitem{pre-LHC}
For a sampling of other pre-LHC analyses, see:
  E.~A.~Baltz and P.~Gondolo,
  JHEP {\bf 0410} (2004) 052
  [arXiv:hep-ph/0407039];
  B.~C.~Allanach and C.~G.~Lester,
  Phys.\ Rev.\  D {\bf 73} (2006) 015013
  [arXiv:hep-ph/0507283];
  R.~R.~de Austri, R.~Trotta and L.~Roszkowski,
  JHEP {\bf 0605} (2006) 002
  [arXiv:hep-ph/0602028];
  R.~Lafaye, T.~Plehn, M.~Rauch and D.~Zerwas,
  Eur.\ Phys.\ J.\  C {\bf 54} (2008) 617
  [arXiv:0709.3985 [hep-ph]];
  S.~Heinemeyer, X.~Miao, S.~Su and G.~Weiglein,
  JHEP {\bf 0808} (2008) 087
  [arXiv:0805.2359 [hep-ph]];
  R.~Trotta, F.~Feroz, M.~P.~Hobson, L.~Roszkowski and R.~Ruiz de Austri,
  JHEP {\bf 0812} (2008) 024
  [arXiv:0809.3792 [hep-ph]];
  P.~Bechtle, K.~Desch, M.~Uhlenbrock and P.~Wienemann,
  Eur.\ Phys.\ J.\  C {\bf 66} (2010) 215
  [arXiv:0907.2589 [hep-ph]].

\bibitem{post-LHC} 
  For a sampling of other post-LHC analyses, see:
 D.~Feldman, K.~Freese, P.~Nath, B.~D.~Nelson and G.~Peim,
  Phys.\ Rev.\ D {\bf 84}, 015007 (2011)
  [arXiv:1102.2548 [hep-ph]];
   B.~C.~Allanach,
  Phys.\ Rev.\ D {\bf 83}, 095019 (2011)
  [arXiv:1102.3149 [hep-ph]];
 S.~Scopel, S.~Choi, N.~Fornengo and A.~Bottino,
  Phys.\ Rev.\ D {\bf 83}, 095016 (2011)
  [arXiv:1102.4033 [hep-ph]];
  P.~Bechtle {\it et al.},
  Phys.\ Rev.\ D {\bf 84}, 011701 (2011)
  [arXiv:1102.4693 [hep-ph]].
 B.~C.~Allanach, T.~J.~Khoo, C.~G.~Lester and S.~L.~Williams,
  JHEP {\bf 1106}, 035 (2011)
  [arXiv:1103.0969 [hep-ph]];
   S.~Akula, N.~Chen, D.~Feldman, M.~Liu, Z.~Liu, P.~Nath and G.~Peim,
  Phys.\ Lett.\ B {\bf 699}, 377 (2011)
  [arXiv:1103.1197 [hep-ph]];
M.~J.~Dolan, D.~Grellscheid, J.~Jaeckel, V.~V.~Khoze and P.~Richardson,
  JHEP {\bf 1106}, 095 (2011)
  [arXiv:1104.0585 [hep-ph]];
  S.~Akula, D.~Feldman, Z.~Liu, P.~Nath and G.~Peim,
  Mod.\ Phys.\ Lett.\ A {\bf 26}, 1521 (2011)
  [arXiv:1103.5061 [hep-ph]];
  M.~Farina, M.~Kadastik, D.~Pappadopulo, J.~Pata, M.~Raidal and A.~Strumia,
  Nucl.\ Phys.\ B {\bf 853}, 607 (2011)
  [arXiv:1104.3572 [hep-ph]];
S.~Profumo,
  Phys.\ Rev.\ D {\bf 84}, 015008 (2011)
  [arXiv:1105.5162 [hep-ph]];
    T.~Li, J.~A.~Maxin, D.~V.~Nanopoulos and J.~W.~Walker,
  arXiv:1106.1165 [hep-ph];
 N.~Bhattacharyya, A.~Choudhury and A.~Datta,
  Phys.\ Rev.\ D {\bf 84}, 095006 (2011)
  [arXiv:1107.1997 [hep-ph]].

\bibitem{LR}  
  A.~Fowlie, A.~Kalinowski, M.~Kazana, L.~Roszkowski and Y.~L.~S.~Tsai,
  Phys.\ Rev.\ D {\bf 85}, 075012 (2012)
  [arXiv:1111.6098 [hep-ph]].

  \bibitem{Barate:2003sz}
  R.~Barate {\it et al.}  [ALEPH, DELPHI, L3, OPAL
                       Collaborations and LEP Working Group for Higgs
                       boson searches],
  Phys.\ Lett.\  B {\bf 565} (2003) 61
  [arXiv:hep-ex/0306033].

\bibitem{Schael:2006cr}
  S.~Schael {\it et al.}  [ALEPH, DELPHI, L3, OPAL
                       Collaborations and LEP Working Group for Higgs
                       boson searches],
  Eur.\ Phys.\ J.\  C {\bf 47} (2006) 547
  [arXiv:hep-ex/0602042].

\bibitem{TevHiggs} {\tt http://tevnphwg.fnal.gov/} and references therein.

\bibitem{ATLASHA}
ATLAS Collaboration, \\
{\tt https://atlas.web.cern.ch/Atlas/GROUPS/}
{\tt PHYSICS/CONFNOTES/ATLAS-CONF-2011-132/}
{\tt ATLAS-CONF-2011-132.pdf}.

\bibitem{CMSHA}
CMS Collaboration, \\
{\tt http://cdsweb.cern.ch/record/1378096/} {\tt files/HIG-11-020-pas.pdf.}

\bibitem{ATLAS+CMS}
CMS Collaboration, \\
{\tt https://cdsweb.cern.ch/ecord/1399607/} {\tt files/HIG-11-023-pas.pdf};  \\
ATLAS Collaboration, \\
{\tt https://cdsweb.cern.ch/record/1399599} {\tt ?ln=enATLAS-CONF-2011-157};\\
L.~Rolandi, on behalf of ATLAS, CMS and the LHC Higgs Combination Group,
{\tt http://indico.in2p3.fr/getFile.py/} {\tt access?contribId=72\&sessionId=19\&resId} {\tt =0\&materialId=slides\&confId=6004}.

\bibitem{Dec13}
F.~Gianotti for the ATLAS Collaboration, G.~Tonelli for the CMS Collaboration,
{\tt http://indico.cern.ch/conferenceDisplay} {\tt .py?confId=164890}.

\bibitem{unstable}
See, for example, 
J.~Ellis, J.~R.~Espinosa, G.~F.~Giudice, A.~Hoecker, A.~Riotto, Phys.\ Lett.\ B {\bf 679} (2009) 369 and references therein.

\bibitem{ER}
J.~Ellis, D.~A.~Ross, Phys.\ Lett.\ B {\bf 506} (2001) 331.

\bibitem{cmssm1}
M.~Drees and M.~M.~Nojiri,
Phys.\ Rev.\ D {\bf 47} (1993) 376 [arXiv:hep-ph/9207234];
G.~L.~Kane, C.~F.~Kolda, L.~Roszkowski and J.~D.~Wells,
  Phys.\ Rev.\  D {\bf 49} (1994) 6173
  [arXiv:hep-ph/9312272];
H.~Baer and M.~Brhlik,
Phys.\ Rev.\ D {\bf 53} (1996) 597 [arXiv:hep-ph/9508321];
  Phys.\ Rev.\  D {\bf 57} (1998) 567
  [arXiv:hep-ph/9706509];
  J.~R.~Ellis, T.~Falk, K.~A.~Olive and M.~Schmitt,
Phys.\ Lett.\ B {\bf 388} (1996) 97
[arXiv:hep-ph/9607292];
Phys.\ Lett.\ B {\bf 413} (1997) 355
[arXiv:hep-ph/9705444];
J.~R.~Ellis, T.~Falk, G.~Ganis, K.~A.~Olive and M.~Schmitt,
Phys.\ Rev.\ D {\bf 58} (1998) 095002
[arXiv:hep-ph/9801445];
V.~D.~Barger and C.~Kao,
Phys.\ Rev.\ D {\bf 57} (1998) 3131
[arXiv:hep-ph/9704403];
J.~R.~Ellis, T.~Falk, G.~Ganis and K.~A.~Olive,
Phys.\ Rev.\ D {\bf 62} (2000) 075010
[arXiv:hep-ph/0004169];
J.~R.~Ellis, T.~Falk, G.~Ganis, K.~A.~Olive and M.~Srednicki,
Phys.\ Lett.\ B {\bf 510} (2001) 236
[arXiv:hep-ph/0102098];
V.~D.~Barger and C.~Kao,
Phys.\ Lett.\ B {\bf 518} (2001) 117
[arXiv:hep-ph/0106189];
L.~Roszkowski, R.~Ruiz de Austri and T.~Nihei,
JHEP {\bf 0108} (2001) 024
[arXiv:hep-ph/0106334];
A.~Djouadi, M.~Drees and J.~L.~Kneur,
JHEP {\bf 0108} (2001) 055
[arXiv:hep-ph/0107316];
U.~Chattopadhyay, A.~Corsetti and P.~Nath,
Phys.\ Rev.\ D {\bf 66} (2002) 035003
[arXiv:hep-ph/0201001];
J.~R.~Ellis, K.~A.~Olive and Y.~Santoso,
New Jour.\ Phys.\  {\bf 4} (2002) 32
[arXiv:hep-ph/0202110];
H.~Baer, C.~Balazs, A.~Belyaev, J.~K.~Mizukoshi, X.~Tata and Y.~Wang,
JHEP {\bf 0207} (2002) 050
[arXiv:hep-ph/0205325];
R.~Arnowitt and B.~Dutta,
arXiv:hep-ph/0211417;
J.~R.~Ellis, K.~A.~Olive, Y.~Santoso and V.~C.~Spanos,
Phys.\ Lett.\ B {\bf 565} (2003) 176
[arXiv:hep-ph/0303043];
H.~Baer and C.~Balazs,
  JCAP {\bf 0305}, 006 (2003)
  [arXiv:hep-ph/0303114];
A.~B.~Lahanas and D.~V.~Nanopoulos,
  Phys.\ Lett.\  B {\bf 568}, 55 (2003)
  [arXiv:hep-ph/0303130];
U.~Chattopadhyay, A.~Corsetti and P.~Nath,
  Phys.\ Rev.\  D {\bf 68}, 035005 (2003)
  [arXiv:hep-ph/0303201];
  C.~Munoz,
  Int.\ J.\ Mod.\ Phys.\  A {\bf 19}, 3093 (2004)
  [arXiv:hep-ph/0309346];
   R.~Arnowitt, B.~Dutta and B.~Hu,
arXiv:hep-ph/0310103.
   J.~Ellis and K.~A.~Olive,
  in ``Particle Dark Matter", ed. G.~Bertone, pp142-163
  [arXiv:1001.3651 [astro-ph.CO]].
  
    \bibitem{enos}
   J.~R.~Ellis, D.~V.~Nanopoulos, K.~A.~Olive and Y.~Santoso,
  Phys.\ Lett.\ B {\bf 633}, 583 (2006)
  [hep-ph/0509331].



    \bibitem{nuhm1}
H.~Baer, A.~Mustafayev, S.~Profumo, A.~Belyaev and X.~Tata,
  Phys.\ Rev.\  D {\bf 71} (2005) 095008
  [arXiv:hep-ph/0412059];
            H.~Baer, A.~Mustafayev, S.~Profumo, A.~Belyaev and X.~Tata,
               {\em JHEP} {\bf 0507} (2005) 065, 
               hep-ph/0504001;
  J.~R.~Ellis, K.~A.~Olive and P.~Sandick,
  Phys.\ Rev.\  D {\bf 78} (2008) 075012
  [arXiv:0805.2343 [hep-ph]].
  
\bibitem{XE100}
E.~Aprile {\it et al.}  [XENON100 Collaboration],
  Phys.\ Rev.\ Lett.\  {\bf 107}, 131302 (2011)
  [arXiv:1104.2549 [astro-ph.CO]].
  
\bibitem{Svenetal}
  S.~Heinemeyer {\it et al.}, 
  JHEP {\bf 0608} (2006) 052
  [arXiv:hep-ph/0604147];
  S.~Heinemeyer, W.~Hollik, A.~M.~Weber and G.~Weiglein,
  JHEP {\bf 0804} (2008) 039
  [arXiv:0710.2972 [hep-ph]].

\bibitem{Allanach:2001kg}
  B.~C.~Allanach,
  Comput.\ Phys.\ Commun.\  {\bf 143} (2002) 305
  [arXiv:hep-ph/0104145].

\bibitem{SuFla}
 G.~Isidori and P.~Paradisi,
  Phys.\ Lett.\ B {\bf 639} (2006) 499
  [arXiv:hep-ph/0605012];
  G.~Isidori, F.~Mescia, P.~Paradisi and D.~Temes,
  Phys.\ Rev.\  D {\bf 75} (2007) 115019
  [arXiv:hep-ph/0703035], and references therein.

\bibitem{SuperIso}
F.~Mahmoudi,
  Comput.\ Phys.\ Commun.\  {\bf 178} (2008) 745
  [arXiv:0710.2067 [hep-ph]]; 
  Comput.\ Phys.\ Commun.\  {\bf 180} (2009) 1579
  [arXiv:0808.3144 [hep-ph]];
  D.~Eriksson, F.~Mahmoudi and O.~Stal,
  JHEP {\bf 0811} (2008) 035
  [arXiv:0808.3551 [hep-ph]];
  A.~Arbey and F.~Mahmoudi,
  Comput.\ Phys.\ Commun.\  {\bf 181} (2010) 1277
  [arXiv:0906.0369 [hep-ph]].

\bibitem{MicroMegas}
  G.~Belanger, F.~Boudjema, A.~Pukhov and A.~Semenov,
  Comput.\ Phys.\ Commun.\  {\bf 176} (2007) 367
  [arXiv:hep-ph/0607059];
  Comput.\ Phys.\ Commun.\  {\bf 149} (2002) 103
  [arXiv:hep-ph/0112278];
  Comput.\ Phys.\ Commun.\  {\bf 174} (2006) 577
  [arXiv:hep-ph/0405253].

\bibitem{SSARD}  Information about this code is available from K.~A.~Olive: it contains important contributions 
from T.~Falk, A.~Ferstl, G.~Ganis, A.~Mustafayev, J.~McDonald, K.~A.~Olive, P.~Sandick, Y.~Santoso and M.~Srednicki. 

\bibitem{SLHA}
P.~Skands {\it et al.},
  JHEP {\bf 0407} (2004) 036
  [arXiv:hep-ph/0311123];
  B.~Allanach {\it et al.},
  Comput.\ Phys.\ Commun.\  {\bf 180} (2009) 8
  [arXiv:0801.0045 [hep-ph]].

\bibitem{ATLASsusy}
G.~Aad {\it et al.}  [ATLAS Collaboration],
  Phys.\ Lett.\ B {\bf 710}, 67 (2012)
  [arXiv:1109.6572 [hep-ex]].

\bibitem{CMSsusy}
 S.~Chatrchyan {\it et al.}  [CMS Collaboration],
  Phys.\ Rev.\ Lett.\  {\bf 107}, 221804 (2011)
  [arXiv:1109.2352 [hep-ex]].
  
\bibitem{CMSbmm}
S.~Chatrchyan {\it et al.}  [CMS Collaboration],
  Phys.\ Rev.\ Lett.\  {\bf 107}, 191802 (2011)
  [arXiv:1107.5834 [hep-ex]].
  
\bibitem{LHCbbmm}
R.~Aaij {\it et al.}  [LHCb Collaboration],
  Phys.\ Lett.\  B {\bf 699} (2011) 330
  [arXiv:1103.2465 [hep-ex]];
  Phys.\ Lett.\ B {\bf 708}, 55 (2012)
  [arXiv:1112.1600 [hep-ex]].
\bibitem{CDFbmm}
T.~Aaltonen {\it et al.}  [CDF Collaboration],
  Phys.\ Rev.\ Lett.\  {\bf 107}, 239903 (2011)
  [Phys.\ Rev.\ Lett.\  {\bf 107}, 191801 (2011)]
  [arXiv:1107.2304 [hep-ex]].
  
\bibitem{AbdusSalam:2011fc}
 S.~S.~AbdusSalam,  {\it et al.},
  Eur.\ Phys.\ J.\ C {\bf 71}, 1835 (2011)
  [arXiv:1109.3859 [hep-ph]].
 
 \bibitem{DarkSusy}
 P.~Gondolo, J.~Edsjo, P.~Ullio, L.~Bergstrom, M.~Schelke and E.~A.~Baltz,
  JCAP {\bf 0407} (2004) 008
  [arXiv:astro-ph/0406204];\\
{\tt http://www.physto.se/~edsjo/darksusy/}.

\bibitem{SuperIsoRelic}
See the last reference in~\cite{SuperIso}.
 
\end{thebibliography}
\end{document}